\documentclass[11pt,a4paper]{article}
\usepackage{jheparxiv}
\usepackage{amsmath}
\usepackage{amsfonts}
\usepackage{amssymb}
\usepackage{latexsym}
\usepackage{mathrsfs}
\usepackage{braket}		
\usepackage{color}
\usepackage{xcolor}
\usepackage{slashed}
\usepackage{twistor}
\usepackage[all]{xy}
\usepackage{tikz-cd}
\usepackage{mathtools}
\usepackage{bm, bbm}
\usepackage{color}
\usepackage{braket}
\usepackage{float}
\usepackage[]{todonotes}
\usepackage{tikz-feynhand}

\makeatletter
\renewcommand{\todo}[2][]{%
    \@todo[caption={#2}, #1]{\begin{spacing}{0.5}#2\end{spacing}}%
} 
\makeatother 
\usepackage{setspace}

\newcommand{\dd}{{\rm d}}
\newcommand{\nn}{\nonumber \\}
\newcommand{\Mpl}{M_{\rm Pl}}
\newcommand{\Amp}{ \mathcal{A} }

\newcommand{\Dbraket}[1]{\Big\langle \!\!\! \Big \langle #1 \Big\rangle \!\!\! \Big \rangle}

\newcommand{\re}{{\rm Re}\,}
\newcommand{\ima}{{\rm Im}\,}

\title{On-Shell Approach to Black Hole Mergers
}
\author[a]{Katsuki Aoki,}
\affiliation[a]{Center for Gravitational Physics and Quantum Information,
Yukawa Institute for Theoretical Physics, Kyoto University, 606-8502, Kyoto, Japan }
\emailAdd{katsuki.aoki@yukawa.kyoto-u.ac.jp}

\author[a]{Andrea Cristofoli,} 
\emailAdd{cristofoli@yukawa.kyoto-u.ac.jp}
\author[c,d]{and Yu-tin Huang}
\affiliation[c]{Department of Physics and Center for Theoretical Physics, National Taiwan University, Taipei
10617, Taiwan}
\affiliation[d]{Physics Division, National Center for Theoretical Sciences, Taipei 10617, Taiwan}
\emailAdd{yutin@phys.ntu.edu.tw}

\abstract{We develop an on-shell approach to study black hole mergers. Since, asymptotically, the initial and final states can be described by point-like spinning particles, we propose a massive three-point amplitude for the merger of two Schwarzschild black holes into a Kerr black hole. This three-point amplitude and the spectral function of the final state are fully determined by kinematics and the model-independent input about the black hole merger which is described by a complete absorption process. Using the Kosower-Maybee-O'Connell (KMOC) formalism, we then reproduce the classical conservation laws for momentum and angular momentum after the merger. As an application, we use the proposed three-point to compute the graviton emission amplitude, from which we extract the merger waveform to all orders in spin but leading in gravitational coupling. Up to sub-subleading order in spin, this matches the classical soft graviton theorem. We conclude with a comparison to black hole perturbation theory, which gives complementary amplitudes which are non-perturbative in the gravitational coupling but to leading order in the extreme mass ratio limit. This also highlights how boundary conditions on a Schwarzschild background can be used to rederive the proposed on-shell amplitudes for merger processes.}

\begin{document}
{\baselineskip0pt
\rightline{\baselineskip16pt\rm\vbox to-20pt{
           \hbox{YITP-24-135}
\vss}}%
}

\maketitle

\section{Introduction}

In recent years, there has been tremendous progress in understanding how to extract classical observables from on-shell scattering amplitudes. The leading motivation is to apply the techniques developed for perturbative amplitude calculations in quantum field theory and gravity to problems involving gravitational waves from binary black holes. These techniques include unitarity cut methods~\cite{Bern:1994zx, Bern:1994cg, Bern:1995db, Bern:1997sc, Britto:2004nc}, integration-by-parts (IBP) reduction~\cite{Laporta:2000dsw, Chetyrkin:1981qh}, the method of differential equations~\cite{Bern:1993kr, Remiddi:1997ny, Gehrmann:1999as, Kotikov:1990kg}, and double copy relations~\cite{Kawai:1985xq, Bern:2008qj, Monteiro:2014cda}. Most of this progress has been made within the framework of the post-Minkowskian (PM) expansion, valid during the inspiral, using either the effective field theory approach~\cite{Cheung:2018wkq, Bern:2019nnu, Bern:2019crd, Cheung:2020gyp, DiVecchia:2020ymx, DiVecchia:2021bdo, Bjerrum-Bohr:2021din} or the effective world-line description~\cite{Kalin:2020mvi, Kalin:2020fhe, Dlapa:2021npj, Mogull:2020sak, Jakobsen:2021smu, Jakobsen:2022psy, Dlapa:2021vgp, Dlapa:2022lmu, Kalin:2022hph, Ivanov:2024sds}. This expansion takes advantage of the separation of scales in the two-body problem when the impact parameter is much larger than the Schwarzschild radius. As a result, the degrees of freedom can be well described by point-like objects exchanging gravitons. Finite-size effects can be incorporated by introducing local operators~\cite{Cheung:2020sdj, Bern:2020uwk,Aoude:2020ygw}.

Given the advent of applying scattering amplitudes to classical gravitational physics, an obvious question arises: can such an approach be applied to describe the black hole merger process? There are two clear challenges to this. First, the merger process, where two black holes coalesce into one, is inherently non-perturbative and lacks a well-defined perturbative expansion parameter. Second, any attempt to use transition amplitudes to describe the process must involve black hole microstates, which come with large degeneracies. The large degeneracy of black hole states suggests that any exclusive observable would be suppressed by a factor of $e^{-S_{BH}/2}$, where $S_{BH}$ represents the black hole entropy \cite{Giddings:2007qq,Giddings:2009gj}.

The non-perturbative nature of the problem can be circumvented by considering an effective description that leverages the fact that the asymptotic states --- isolated black holes --- are well-defined. The non-perturbative dynamics are local and can be encapsulated into effective operators. This approach is akin to the chiral Lagrangian in its treatment of pion scattering. In this case, the effective description primarily involves a three-point coupling between two scalar particles and a spinning state, providing a simplified framework for capturing the essential dynamics of black hole mergers
\begin{equation}
\mathcal{A}(12 X^\ell)=(p_1{-}p_2)_{\mu_1}\cdots(p_1{-}p_2)_{\mu_\ell}\varepsilon_{X}^{\mu_1\cdots \mu_\ell}\,.
\label{3pt_p}
\end{equation}
In this approach, the two incoming scalars represent two Schwarzschild black holes, which, at sufficiently large distances, are well described as point-like particles. The final state, denoted by $X$, is a minimally coupled spin-$\ell$ state, representing the resulting Kerr black hole~\cite{Guevara:2018wpp, Chung:2018kqs, Arkani-Hamed:2019ymq,Aoude:2020onz}. Importantly, the amplitude for this process is entirely determined by kinematics, making it a powerful tool for describing the black hole merger in terms of scattering amplitudes.\footnote{Note that on-shell $2p_1\cdot p_2=m_X^2-m_1^2-m_2^2$ so that it's dependence can be absorbed into the overall dimensionful coupling.} This allows for a simplified yet effective description of the complex dynamics involved in the merger process.

There are two things amiss here. One is that the final state should be a Kerr black hole carrying \textit{classical} angular momentum, while the other is the spectrum for the final state $X$. The former can be straightforwardly taken care of by considering coherent spin states~\cite{Aoude:2021oqj}, where one superimposes states with definite spin, such that the expectation value of the spin operator matches with the classical spin. The second missing information is encapsulated in the unknown spectral density of the final black hole states. Embracing our effective field theory mantle, we will fix it through matching conditions. The candidate observable is the absorption cross-section, which has been recently used to compute absorptive effects from three-point amplitudes~\cite{Aoude:2023fdm, Jones:2023ugm, Chen:2023qzo}. The total orbital angular momentum of the two Schwarzschild black holes is determined by their momenta and impact parameter. When this angular momentum falls below a certain threshold, the black holes will merge, and the total cross-section becomes fully described by the absorption cross-section. By matching this to the complete absorption cross-section, the spectral function can be universally determined in the classical limit. The exact threshold for the angular momentum, which governs whether the merger occurs, is model-dependent and must be identified either through phenomenological methods or numerical simulations. This provides a seamless connection between the scattering amplitude approach and the physical process of black hole mergers.

We then turn to the issue of extracting observables for black hole merger processes using our on-shell approach. While exclusive observables are not suitable for such processes, inclusive observables --- such as gravitational waveforms --- are readily accessible. Waveforms associated with the merger process have already been incorporated into the effective-one-body (EOB) approach by matching inputs from the inspiral phase to the quasinormal modes of the ringdown~\cite{Buonanno:1998gg, Buonanno:2000ef} (see~\cite{Damour:2012mv} for a review). For our purposes, it is more natural to consider the Kosower, Maybee, O'Connell (KMOC) formalism~\cite{Kosower:2018adc} which has provided remarkable insights on the relations between classical observables and on-shell amplitudes \cite{Maybee:2019jus, Monteiro:2021ztt, Monteiro:2020plf, Luna:2023uwd, Damgaard:2023ttc, Damgaard:2023vnx,Cristofoli:2021vyo,Cristofoli:2021jas}. Indeed, we will show that, starting from the three-point amplitude, momentum and angular momentum conservations are recovered for merger processes when computing in-in expectation values within the KMOC framework. Equipped with the proposed three-point amplitude, this will allow us to compute salient features of the waveform using the KMOC formalism~\cite{Cristofoli:2021vyo}
\begin{align}
h_{\mu\nu} = \sum_{\sigma} \kappa \int \dd \Phi(k) \varepsilon_{\mu\nu}^{-\sigma} e^{-ik\cdot x} i W^{\sigma}(p_i, k, b_i)
    +{\rm c.c.}
\end{align}
where at the leading order, the spectral waveform is given by
\begin{align}
    iW^{\sigma} 
&\,= \Dbraket{ \int_{X}\,\int \prod_{i=1,2} \hat{\dd}^4 q_i \hat{\delta}(2p_i \cdot q_i) e^{-ib_i \cdot q_i}  
\underbrace{\bra{p_1+q_1; p_2+q_2}T^{\dagger}\ket{p_X,\alpha}}_{\text{massive 3pt}}
\, \underbrace{\bra{p_X,\tilde{\alpha};k^{\sigma}} T \ket{p_1; p_2}}_{\text{4pt with one graviton}} }
    \,.
    \label{W_intro}
\end{align}
We will explain our notation later so there is no need to look at details now. The point of this formula is that the waveform for the merger is computed by the three-point amplitude described previously and the four-point amplitude with one additional graviton emission. We construct the required four-point amplitude via on-shell gluing at leading order in the gravitational coupling with no assumption on the mass ratio, thus including all orders in the spin of the final black hole. Remarkably the amplitude comes in a factorized form
\begin{align}
\underbrace{\Amp(p_X,\tilde{\alpha}; k^{\pm}|p_1 ; p_2)}_{\text{4pt with one graviton}} &=\underbrace{\Amp(p_{12},\tilde{\alpha}|p_1; p_2)}_{\text{massive 3pt}} S^{\pm}_{\alpha}
\,,
\end{align}
where the precise form of the factor $S^{\pm}_{\alpha}$ will be given in eq.~(\ref{Splus}). Using eq.~\eqref{W_intro}, the waveform is simply given by the classical limit of $S^{\pm}_{\alpha}$ with kinematics fixed by our proposed three-point. We will then verify up to the quadratic order in spin that it reproduces the universal form of the {\it classical} version of the sub-subleading soft graviton operator~\cite{Cachazo:2014fwa, Pasterski:2015tva, Laddha:2019yaj, Laddha:2017ygw, Laddha:2018rle}, thus providing an on-shell proof of the equivalence between gravitational (spin) memory and soft theorems at sub-subleading order \cite{Laddha:2017ygw,Freidel:2021dfs}. 

We conclude by considering a similar setup in black hole perturbation theory, where we study the propagation of a massive scalar in a black hole background. Once the WKB limit is considered, this scenario can be viewed as the large mass ratio limit of the black hole merger problem, though it remains non-perturbative in the gravitational coupling. This can give a complementary discussion to the one mentioned above while giving a clear comparison with the conventional classical approach to the black hole merger. In doing so, we first show how different mode functions of the field propagating on the black hole spacetime are related to on-shell amplitudes for elastic (scattering) and inelastic (merger) processes. We also provide the critical angular momentum for the merger, reproducing the result from~\cite{Unruh:1976fm}.
Next, we consider the emission process during the merger using the KMOC formalism for large mass ratios. For simplicity, we use a massless scalar field that couples to the probe massive scalar to mimic the graviton emission process. The result is directly compared to the waveform computed by solving the classical equations of motion, demonstrating complete equivalence. This verifies that on-shell amplitudes can indeed be used to compute black hole mergers and the associated waveforms.
\\

\textbf{Conventions:} Unless stated otherwise we work in natural units $c=\hbar=1$ with the gravitational coupling $\kappa^2=32\pi G$. We will use the 
mostly minus signature of the metric \\ $\eta_{\mu\nu}={\rm diag}[+1, -1, -1, -1]$. The on-shell phase-space integral is denoted by
\begin{align}
    \int_p :=\int \dd \Phi(p)\,, \qquad \dd \Phi(p):=\hat{\dd}^4p \: \hat{\delta}^{(+)}(p^2-m^2)
\end{align}
with the $(2\pi)$ normalised measure and delta functions
\begin{align}
    \hat{\dd}^n p:= \frac{\dd^n p}{(2\pi)^n}\,, \qquad \hat{\delta}^{(n)}(p):=(2\pi)^n \delta^{(n)}(p)
    \,.
\end{align}
The process of scattering amplitudes is specified by either the subscript or the argument. In the latter case, we will use the notation that in and out states are separated by the vertical bar: the left (right) labels are the ones for out (in) states by following the bra-ket notation. The labels of different particles in the in/out states are further separated by semicolons. For instance, the 3-point amplitude for two incoming spin-0 particles and one outgoing particle with the angular momentum $(\ell,m)$ is $\Amp(p_X,\ell,m|p_1;p_2)$ being subject to the conservation law $p_1+p_2=p_X$. We will also use the spinor-helicity notation with the spinors being denoted by $\lambda, \tilde{\lambda}$ or the bra-ket notation by following~\cite{Arkani-Hamed:2017jhn}.

\section{Black hole formation in S-matrix}
\label{sec:BH_in_S-matrix}
We consider a scattering problem of two Schwarzschild BHs of masses $m_1$ and $m_2$. The system has a super-Planckian energy $E\gg \Mpl$ because the objects are macroscopic. In such a super-Planckian scattering, a new BH state can be formed together with the GW emission as it is classically predicted by a coalescence of two BHs. This process is analogous to nuclear fusion reactions and may be systematically written as
\begin{align}
{\rm BH} + {\rm BH} \to {\rm BH} + {\rm gravitons}
\,.
\end{align}
We investigate the S-matrix description of such classical BH formulations and compute the associated observables by the KMOC formalism. Super-Planckian scatterings have been investigated in~\cite{Giddings:2007qq,Giddings:2009gj}. However, our focus is a scattering of macroscopic objects rather than elementary particles. This would simplify the construction of the S-matrix. Let's explore the S-matrix for classical black hole formations. 

\subsection{Black hole formation in 2-to-2 scattering}

We start with the partial wave expansion of the $m_1m_2 \to m_1 m_2$ scattering amplitude
 \begin{align}
\Amp_{12\to 12}(s,\cos \theta)=16\pi \sum_{\ell=0}^{\infty}(2\ell +1) a_{\ell}(s) P_{\ell}(\cos \theta)
\,.
\label{12->12}
\end{align}
The partial wave amplitude can be written in terms of the phase shift $\delta_{\ell}(s)$ and the elasticity parameter $\eta_{\ell}(s)$:
\begin{align}
a_{\ell}(s)
&= \frac{E}{2P} \frac{\eta_{\ell} e^{2i\delta_{\ell}}-1}{2i}
\nn
&= \frac{s}{2\sqrt{(p_1 \cdot p_2)^2-m_1^2m_2^2}} \frac{\eta_{\ell} e^{2i\delta_{\ell}}-1}{2i}\,.
\end{align}
Since in the partial wave basis the S-matrix is given as $S_{\ell}=\eta_{\ell}e^{2i\delta_{\ell}}$, allowing for inelastic scattering $|S_{\ell}|^2\leq 1$ leads to  
\begin{align}
0\leq \eta_{\ell} \leq 1\,.
\end{align}
Here, $E$ and $P$ are the centre-of-mass energy and momentum
\begin{align}
    E=\sqrt{s}\,, \quad P=\sqrt{\frac{(p_1 \cdot p_2)^2-m_1^2m_2^2}{s}} = \sqrt{ \frac{\lambda(s,m_1^2, m_2^2)}{4s} } \,,
\end{align}
where $\lambda(x,y,z):=x^2+y^2+z^2-2xy-2yz-2zx$ is the K\"{a}ll\'{e}n function.\footnote{In literature, $\gamma:=p_1\cdot p_2/m_1m_2$ is often used by which we have $P=\frac{m_1m_2}{\sqrt{s}}\sqrt{\gamma^2-1}$.} The case $\eta_{\ell}=1$ corresponds to the elastic process, i.e.,~conservative dynamics, $0\leq \eta_{\ell}<1$ means the existence of inelastic channels like graviton emissions, and $\eta_{\ell}=0$ is complete absorption.

Let us discuss a general expectation of the behaviour of $\eta_{\ell}(s)$ by referring to the classical two-body problem. In the classical limit, $\ell$ may be identified with the classical orbital angular momentum of the system. For a given centre-of-mass energy, the BHs will scatter and then back to infinity for a large angular momentum. The distance of the closest approach of the two BHs will decrease as $\ell$ decreases, and accordingly, more radiation will be emitted. The elasticity $\eta_{\ell}$ thus also decreases for a small $\ell$, but still remains finite as long as we are considering the scattering regime. When $\ell$ falls below a critical value, which would be of the order of $Gm_1m_2$, the BHs will collide, forming a new BH state and emitting GWs. Classically, the system never returns to the original two-body system once the new BH is formed. Hence, BH mergers should be described by the complete absorption $\eta_{\ell}=0$ in the 2-to-2 scattering.\footnote{At the quantum level, it might be possible to recover the initial state by Hawking radiation if no information is lost. The elasticity parameter may be given by $\eta_{\ell}=e^{-S_{\rm BH}/2}$ where $S_{\rm BH}$ is the Bekenstein-Hawking entropy~\cite{Giddings:2007qq,Giddings:2009gj}. The deviation from the complete absorption is negligible for classical BHs, $S_{\rm BH}\gg 1$.}

For simplicity, we neglect the GW emission in this section and come back to the issue of radiation later. Then, in the classical limit, the elasticity parameter should be given by
\begin{align}
\eta_{\ell}=
\begin{cases}
1\,, & \ell > L_c\,, \\
0\,, & \ell \leq L_c\,, 
\end{cases}
\end{align} 
where $L_c(s)$, which depends on the centre-of-mass energy, is a critical angular momentum of forming a new BH. The critical value can be determined to match with the classical geodesic motion when one of the BHs is a test particle. It is also possible to compute the absorption cross-section of BH at the quantum level and then take the classical limit~\cite{Unruh:1976fm}. They will be discussed in Sec.~\ref{sec:BHspacetime}. However, this is a model {\it dependent} part requiring a test particle approximation of one BH or in the general case a reference to numerical relativity while our analysis here is model {\it independent}. In fact, as we will see, the detailed value of $L_c$ is not important for our discussion and we just need to remember that the typical size is given by $L_c \sim Gm_1m_2$. The total cross-section is then
\begin{align}
\sigma^{\rm tot}_{\ell}=
\begin{cases}
\sigma_{\ell}^{\rm el}\,, & \ell > L_c \,, \\
\sigma_{\ell}^{\rm el}+\sigma_{\ell}^{\rm abs}\,, & \ell \leq L_c
\,,
\end{cases}
\end{align}
where $\sigma_{\ell}^{\rm el} $ and $\sigma_{\ell}^{\rm abs}$ are the elastic and absorption cross-sections of partial waves, respectively. As is well known (e.g.,~\cite{Schwartz:2014sze}), the partial wave cross-sections are computed by
\begin{align}
\sigma^{\rm el}_{\ell}=\frac{16 \pi}{E^2} (2 \ell+1)\left|a_{\ell}\right|^2\,, \quad
\sigma^{\rm tot}_{\ell}=\frac{8\pi}{EP}(2\ell +1)\ima a_{\ell}
\,,
\end{align}
and the complete absorption $\eta_{\ell}=0$ results in
\begin{align} 
\sigma^{\rm el}_{\ell}=\sigma^{\rm abs}_{\ell}=\frac{1}{2}\sigma^{\rm tot}_{\ell} = \frac{\pi(2\ell + 1)}{P^2}\,,
\qquad (\ell \leq L_c) \,.
\label{cross-sections}
\end{align}
Since we have neglected the radiation, the absorption cross-section should only come from the formation of (spinning) BHs. We may describe such BH states by continuous degrees of freedom, denoted by $X$, with the spectral density $\rho_{\ell}(m_X^2)$, as with the gravitational wave absorption of BHs~\cite{Jones:2023ugm, Aoude:2023fdm, Chen:2023qzo}. Later on in section~\ref{sec:BHspacetime} we will identify these modes within black hole perturbation theory. Therefore, the absorption cross-section is given by
\begin{align}
\sigma^{\rm abs}_{\ell}&=\frac{1}{4EP} \sum_{m=-\ell}^{\ell} \int_{(m_1+m_2)^2}^{+\infty} \dd m_X^2 \dd \Phi(X) \rho_{\ell}(m_X^2) |\Amp(p_X, \ell, m|p_1; p_2)|^2 \hat{\delta}^{(4)}(p_1+p_2-p_X)
\nn
&=\frac{\pi}{2E P}\sum_{m=-\ell}^{\ell} \int_{(m_1+m_2)^2}^{+\infty} \dd m_X^2 \delta(s-m_X^2) \rho_{\ell}(m_X^2 ) |\Amp(p_X, \ell, m|p_1; p_2)|^2 
\,,
\label{abs_cross-section}
\end{align}
where $\dd \Phi(X) =\hat{\dd}^4p_X \hat{\delta}^{(+)}(p_X^2-m_X^2)$ is the Lorentz-invariant phase-space measure of the $X$ state and $\Amp(p_X, \ell, m|p_1; p_2)$ is the three-point amplitude of two incoming Schwarzschild BHs and an outgoing BH with the quantum number $(\ell, m)$. 

Eqs.~\eqref{cross-sections} and \eqref{abs_cross-section} give a matching condition of the three-point amplitude under neglecting radiation. In the following, we determine the concrete expression of the three-point amplitude by using the on-shell method and the matching condition.

\subsection{3-point amplitude for black hole merger: definite spin}
\label{sec:3pt_ell}

By looking at the quantum numbers of the external states, the amplitude $\Amp(p_X, \ell, m|p_1; p_2)$ is nothing but the three-point amplitude of two incoming massive spin-0 particles with momenta $p_1,p_2$ and one outgoing massive spin-$\ell$ particle with the momentum $p_X$. Using the massive spinor-helicity formalism~\cite{Arkani-Hamed:2017jhn}, this three-point amplitude is given by
\begin{align}
    \Amp^{I_1\cdots I_{2\ell }} (p_X, \ell |p_1 ; p_2 ) 
&=g_{\ell} \bra{X^{\{ I_1}}p_1 p_2 \ket{X^{I_2}}  \times \cdots \times \bra{X^{I_{2\ell-1}}}p_1 p_2 \ket{X^{I_{2\ell}\} }}
\nn 
&= g_{\ell} \bra{\bm{X}}p_1 p_2 \ket{\bm{X}}^{\ell}
\,,
\label{M3pt}
\end{align}
where $I_i$ are the SU(2) little group indices.\footnote{This can be recast into the familiar polarization tensor form $(\varepsilon\cdot(p_1-p_2))^\ell$~i.e.,~\eqref{3pt_p}, by identifying $\varepsilon^{I_1I_2}_{\alpha \dot{\alpha}}=\frac{\lambda^{(I_1}_\alpha\tilde{\lambda}^{I_2)}_{\dot{\alpha}}}{m}$.} We use the bold notation of \cite{Arkani-Hamed:2017jhn} by suppressing the little group indices. We recall that the mass of each particle is $p_1^2=m_1^2, p_2^2=m_2^2$ and $p_X^2=m_X^2$ which all are different, in general. 
For reference, the explicit contraction of the SL(2,$\mathbb{C}$) indices is
\begin{align}
\bra{X^{I}}p_1 p_2 \ket{X^{J}}= \bra{X^{I}}^{\alpha} (p_1)_{\alpha\dot{\alpha}} (p_2)^{\dot{\alpha}\beta} \ket{X^{J} }_{\beta} 
\end{align}
where $p_{\alpha\dot{\alpha}}=p_{\mu}(\sigma^{\mu})_{\alpha\dot{\alpha}}, p^{\dot{\alpha}\alpha}=p_{\mu}(\bar{\sigma}^{\mu})^{\dot{\alpha}\alpha}$ with the Pauli matrices $\sigma^{\mu}=(1,\sigma^1,\sigma^2,\sigma^3)$ and $\bar{\sigma}^{\mu}=(1,-\sigma^1,-\sigma^2,-\sigma^3)$. The SL(2,$\mathbb{C}$) indices and the SU(2) little group indices are raised and lowered by the anti-symmetric tensors with $\epsilon^{12}=-\epsilon_{12}=+1$; for example, $\lambda_{\alpha}=\epsilon_{\alpha\beta}\lambda^{\beta}, \lambda^{\alpha}=\epsilon^{\alpha\beta}\lambda_{\beta}$. 
We use the following realisation of the massive spinors by boosting the particle from the rest frame by the rapidity $\lambda$:
\begin{align}
\ket{X^1}_{\alpha} &=
e^{-\frac{\lambda}{2}(\vec{n}\cdot \vec{\sigma}) }
\begin{pmatrix}
\sqrt{m_X} \\ 0
\end{pmatrix}
, \quad
\ket{X^2}_{\alpha} =
e^{-\frac{\lambda}{2}(\vec{n}\cdot \vec{\sigma}) }
\begin{pmatrix}
0 \\ \sqrt{m_X} 
\end{pmatrix}
, \label{massive_spinor1}
\\
|X^1]^{\dot{\alpha}} &=
e^{\frac{\lambda}{2}(\vec{n}\cdot \vec{\sigma}) }
\begin{pmatrix}
\sqrt{m_X} \\ 0
\end{pmatrix}
, \quad
|X^2]^{\dot{\alpha}} =
e^{\frac{\lambda}{2}(\vec{n}\cdot \vec{\sigma}) }
\begin{pmatrix}
0 \\ \sqrt{m_X} 
\end{pmatrix}
,
\end{align}
where $\vec{n}$ is the unit three-dimensional vector and the boosted four-momentum is $p_X^{\mu}=(\mu \cosh \lambda, \mu \sinh \lambda\, \vec{n})$. The amplitude of two outgoing spin-0 and one incoming spin-$\ell$ is given by the one with lower little group indices with the following relation for real kinematics:
\begin{align}
\Amp_{I_1\cdots I_{2\ell}} (p_1 ; p_2|p_X, \ell) = [ \Amp^{I_1\cdots I_{2\ell}} (p_X, \ell |p_1 ; p_2)]^*
\,.
\end{align}

Let us explicitly write the components of the amplitudes for reference. Considering the centre-of-mass frame of incoming particles, which is the rest frame of $p_X$,
\begin{align}
    p_1^{\mu}&=(E_1, P \sin\theta \cos \phi, P \sin\theta \sin\phi, P \cos \theta) \,,\\
    p_2^{\mu}&=(E_2, -P \sin\theta \cos \phi, -P \sin\theta \sin\phi, -P \cos \theta) \,, \\
   p_X^{\mu}&=(m_X, 0, 0, 0)\,,
\end{align}
the components of the 3-point amplitude are given by
\begin{align}
\Amp^{\overbrace{11\cdots 1}^{\ell-m} \overbrace{22\cdots 2}^{\ell + m}} 
 &= g_{\ell} N_{\ell, m} [m_X^2 \lambda(m_1^2,m_2^2,m_X^2)]^{\ell/2}  (-1)^{\ell} Y^*_{\ell, m}(\theta ,\phi)
 \,, 
 \label{3pt_up}
\end{align}
where 
\begin{align}
    N_{\ell,m}:=\frac{\ell!}{(2\ell)!} \sqrt{\frac{4\pi}{2\ell +1} (\ell -m)! (\ell + m)!}
    \,,
\end{align}
and our convention of the spherical harmonics is
\begin{align}
    Y_{\ell, m}:= (-1)^{m}\sqrt{\frac{2\ell+1}{4\pi} \frac{(\ell-m)!}{(\ell +m)!}}P_{\ell}^{m}(\cos\theta)e^{i m\phi}
    \,.
\end{align}
Hence, the massive three-point is nothing but spherical harmonics (see also~\cite{Aoude:2023fdm,Chen:2023qzo}). Nonetheless, the spinor-helicity notation has practical advantages: it is manifestly covariant and is given by a simple tensor product. In particular, the latter is important to take the classical spin limit, as we will see in subsequent sections.

Having understood the kinematic dependence, we now fix the coupling constants $g_{\ell}$ by matching with the complete absorption~\eqref{cross-sections} and \eqref{abs_cross-section} for $\ell<L_c$:
\begin{align}
\frac{4m_X^2(2\ell+1)}{\lambda^{1/2}(m_1^2,m_2^2,m_X^2)}&=\rho_{\ell}(m_X^2)\Amp_{I_1\cdots I_{2\ell}} (p_1 ; p_2|p_X, \ell) \Amp^{I_1\cdots I_{2\ell }} (p_X, s|p_1 ; p_2 )
\nn
&=\rho_{\ell}(m_X^2)|g_{\ell}|^2 \frac{(\ell!)^2}{(2\ell)!} [m_X^2 \lambda(m_1^2,m_2^2,m_X^2)]^{\ell} 
\,.
\end{align}
The summation over the magnetic quantum number is replaced with the contraction of the little group indices.
Note that the spectral density and the coupling constant cannot be uniquely fixed from this matching condition alone. However, it is not problematic for us because they will always appear in the combination $\rho_{\ell} |g_{\ell}|^2$ in observables. Instead, we may use this non-uniqueness to choose the forms of $\rho_{\ell}$ and $g_{\ell}$ for our convenience. It will turn out shortly that a convenient choice is\footnote{Here we choose the normalization such that the three-point amplitude has mass dimension one to make the phase space integral, i.e. $\int dm^2_X \rho_\ell$, dimensionless and the three-point in the coherent spin will be given by the simple form \eqref{3pt_alpha}. Then the gluing of the three-point  integrated over the phases space, leads to a dimensionless four-point amplitude. }
\begin{align}
\rho_{\ell}&=\frac{4(2\ell+1)}{\lambda^{1/2}(m_1^2,m_2^2,m_X^2)}\theta(1-\ell/L_c)
\,, 
\label{rhoell}\\
g_{\ell} &= m_X \frac{\sqrt{(2\ell)!}}{\ell!} [m_X^2 \lambda(m_1^2,m_2^2,m_X^2)]^{-\ell/2}
\,.
\label{gell}
\end{align}
We have multiplied a Heaviside function to guarantee that the states exist only in small $\ell$ classically: a classical BH should form only when the initial angular momentum is below the critical value $L_c$. Notice also that the function $\rho_{\ell}$, while determining the distributions of final states, appears to depend on information related to initial states, such as $m_{1}$ and $m_2$. However, this dependence is artificial to simplify the three-point coupling $g_{\ell}$ and can be removed by redefining $\rho_{\ell}$ up to an overall function. The three-point amplitude of the BHs is therefore
\begin{align}
\Amp^{I_1\cdots I_{2\ell }} (p_X, \ell|p_1 ; p_2 ) = m_X \frac{\sqrt{(2\ell)!}}{\ell !} \left( \frac{\bra{\bm X}p_1  p_2 \ket{\bm X}}{m_X \lambda^{1/2}(m_1^2, m_2^2, m_X^2)} \right)^{\ell}
\,.
\label{3pt_ell}
\end{align}

\subsection{3-point amplitude for black hole merger: coherent spin}
So far, the states $X$ are regarded as ``quantum'' spinning BHs in the sense that they are labelled by discrete quantum numbers $(\ell, m)$. To make contact with ``classical'' spinning BHs, on the other hand, it would be more appropriate to label them with continuous numbers.

We adopt the coherent spin formalism introduced by~\cite{Aoude:2021oqj}. We begin with introducing the 2-dimensional harmonic oscillator satisfying the relation
\begin{align}
\left[ \hat{a}^I, \hat{a}^{\dagger}_J \right] = \delta^I_J
\end{align}
where the indices $I,J=1,2$ are identified with the massive little group SU(2). The spin-$\ell$ state can be constructed from the spin-zero state $\ket{0}$ by acting the creation operators
\begin{align}
\ket{\ell, \{I_1 \cdots I_{2\ell} \} }:=\frac{1}{\sqrt{(2\ell)!}} \hat{a}^{\dagger}_{I_1} \cdots \hat{a}^{\dagger}_{I_{2\ell}} \ket{0}
\,.
\end{align}
The coherent spin state is then defined by a weighted sum of these spin states as
\begin{align}
\ket{\alpha}&:=e^{-\frac{1}{2}\tilde{\alpha}_J \alpha^J} e^{\alpha^I \hat{a}^{\dagger}_I } \ket{0}
\nn
&\, =e^{-\frac{1}{2}\|\alpha \|^2 } \sum_{2\ell=0}^{\infty} \frac{1}{\sqrt{(2\ell)!}} \alpha^{I_1} \cdots \alpha^{I_{2\ell}}\ket{\ell, \{I_1 \cdots I_{2\ell} \} }
\,,
\label{def_alpha}
\end{align}
where $\alpha^I$ is the SU(2) variable and $\tilde{\alpha}_I$ is its complex conjugate. The completeness relation for the spin reads
\begin{align}
\mathbbm{1}_{\rm spin} = \sum_{2\ell=0}^{\infty}\ket{\ell, \{I_1 \cdots I_{2\ell} \}} \bra{\ell, \{I_1 \cdots I_{2\ell} \}} = \int \frac{\dd^2\alpha \dd^2 \tilde{\alpha}}{\pi^2}\ket{\alpha} \bra{\tilde{\alpha}}
\,.
\end{align}

We temporarily recover the $\hbar$ dependence to see the classical spin limit. The covariant spin operator for the particle with the momentum $p_X$, namely the Pauli-Lubanski pseudovector, is defined by
\begin{align}
\mathbb{S}_X^{\mu} :=  \frac{\hbar}{2} \hat{a}^{\dagger}_I [\sigma^{\mu}_X]^I{}_J \hat{a}^J
\,, \label{spin_op}
\end{align}
using the momentum-dependent $\sigma$-matrices
\begin{align}
[\sigma^{\mu}_X]^I{}_J = \frac{1}{2m_X}\left( \bra{X^I}\sigma^{\mu}|X_J] + [X^I|\bar{\sigma}^{\mu}\ket{X_J} \right)
\,.
\end{align}
The expectation value is
\begin{align}
S_X^{\mu}= \bra{p_X, \tilde{\alpha}} \mathbb{S}^{\mu}_X \ket{p_X, \alpha}  =  \frac{\hbar}{2} \tilde{\alpha}_I [\sigma^{\mu}_X]^I{}_J \alpha^J
\,.
\end{align}
Hence, we must scale $\alpha^I \sim \tilde{\alpha}_I \sim \hbar^{-1/2}$ for a particle to have a classical spin $S_X^{\mu} = \mathcal{O}(\hbar^0)$. As discussed in~\cite{Aoude:2021oqj}, the uncertainty of the spin is minimised in the coherent spin state. In particular, the variance vanishes in the classical limit $\hbar \to 0$, allowing for a well-defined classical spin limit.

Let us construct the three-point amplitude in terms of the coherent spin state. The coupling constant \eqref{gell} and the tensor product structure of the 3-point \eqref{M3pt} enable us to resum the weighted sum into the exponential function
\begin{align}
\Amp(p_X, \tilde{\alpha}|p_1,p_2) &= e^{-\frac{1}{2}\|\alpha\|^2} \sum_{2\ell=0}^{\infty} \frac{1} {\sqrt{(2\ell)!}} \tilde{\alpha}_{I_1}\cdots \tilde{\alpha}_{I_{2\ell}} \Amp^{I_1\cdots I_{2\ell}} (p_X, \ell|p_1; p_2) 
\nn
&=m_X e^{-\frac{1}{2}\|\alpha\|^2}   \sum_{\ell=0}^{\infty}  \frac{1}{\ell!} z^{\ell}
\nn
&=m_X e^{-\frac{1}{2}\|\alpha\|^2 +z} 
\,,
\label{3pt_alpha}
\end{align}
where
\begin{align}
z(p_1,p_2):=\frac{\tilde{\alpha}_I\bra{p_{12}^{I}}p_1 p_2 \ket{p_{12}^{J}} \tilde{\alpha}_J}{ m_X \lambda^{1/2}(m_1^2,m_2^2,m_X^2)}
\,,
\label{zdef}
\end{align}
and $\ket{p_{12}^{J}}$ is the massive spinor associated with momentum $p_X=p_1{+}p_2$. It is interesting to notice that the three-point amplitude generically vanishes in the classical limit due to the exponential suppression of the normalisation factor $e^{-\|\alpha \|^2/2}$ unless it is cancelled by the exponentiation of the spinning amplitudes. This is reminiscent of what happens in the three-point amplitude of Kerr BHs and a graviton~\cite{Aoude:2021oqj}. 
In the next subsection, around eq.~\eqref{eq:cancel}, we will elaborate on this cancellation when we study the classical observables.

All left is to transform the spectral density in the basis labelled by coherent spin states. The relation is given starting from the completeness relation for spin and mass-changing states
\begin{equation}
\sum_{2\ell=0}^{\infty}\int \d m_X^2  \ket{\ell, \{I_1 \cdots I_{2\ell} \}} \bra{\ell, \{I_1 \cdots I_{2\ell} \}} \rho_{\ell}= \int \frac{\d^2\alpha \d^2 \tilde{\alpha}}{\pi^2}\d m_X^2 \ket{\alpha} \bra{\tilde{\alpha}}\rho_{\alpha}
\,.
\end{equation}
The contraction over a generic spin-$\ell$ state is
\begin{equation}
 (2\ell)!\delta_{J_1}^{\left\{I_1\right.} \cdots \delta_{J_{2 \ell}}^{\left.I_{2 \ell}\right\}}\rho_{\ell}=  \int \frac{\d^2\alpha \d^2 \tilde{\alpha}}{\pi^2}\rho_{\alpha} e^{-||\alpha||} \alpha^{I_1} \cdots \alpha^{I_{2\ell}} \tilde{\alpha}_{J_1} \cdots \tilde{\alpha}_{J_{2\ell}}
 \,.
\end{equation}
Because of the invariance under the little group, the spectral density in the coherent state basis must be a function of the norm only, $\rho_{\alpha}=\rho_{\alpha}(||\alpha||^2)$. We can then perform a major part of the integral while also removing the tensorial structure
\begin{equation}
 \rho_{\ell}=\frac{1}{\Gamma(2\ell+2)}\int_{0}^{+\infty} \d r \rho_{\alpha}(r)\: e^{-r} r^{2\ell+1} 
\end{equation}
where $r=||\alpha||^2$. One can recognize that this is nothing but the Mellin transform. We can complexify the angular momentum $\ell$ and obtain $\rho_{\alpha}$ by the inverse Mellin transform if $\Gamma(2\ell+2) \rho_{\ell}$ is analytic in the strip $a<\re \ell < b$ and vanishes sufficiently fast in $\ima \ell \to \pm \infty$. In our case, however, our classical spectral function $\rho_{\ell}$ has the Heaviside function \eqref{rhoell} to truncate $\ell$ up to the critical value $L_c$, so the inverse Mellin transform cannot be applied. Note that the Heaviside function should be an approximation in the classical limit because there is no sharp cutoff of the absorption at the quantum level. It may be possible to find a smooth function $\rho_{\ell}$ at the quantum level and then apply the inverse transform. Alternatively, the classical limit of $\rho_{\alpha}$ can be directly infered without passing through quantum theory as follows. One expects that a truncation similar to $\ell$ would appear also in the coherent spin state. In fact, by multiplying an appropriate factor to reproduce the factor $2\ell+1 \approx 2\ell$, the truncated integral gives
\begin{align}
 \frac{1}{\Gamma(2\ell+2)} \int^{a_c}_0 \d r  e^{-r}r^{2\ell +2}
=\frac{\gamma(2\ell+3, a_c)}{\Gamma(2\ell+2)}
&\approx 2\ell \theta(1-2\ell/\text{a}_c)
\,,
\end{align}
in the large $\ell, a_c=\mathcal{O}(\hbar^{-1})$ limit. Here, $\gamma(a,z)$ is the incomplete gamma function and its ratio to the gamma function is asymptotic to the complementary error function~\cite{temme1975uniform,temme1979asymptotic}, which we may approximate by the Heaviside function. Therefore, the truncation in $\alpha$ should be equivalent to the truncation in $\ell$ in the classical limit.\footnote{Precisely speaking, we have only shown that the truncation in $\alpha$ implies the truncation in $\ell$ in the classical limit but not the converse.} As a result, the classical spectral density in the coherent spin basis can be written as
\begin{align}
\rho_{\alpha}&= \frac{4 \|\alpha \|^2 }{\lambda^{1/2}(m_1^2,m_2^2, m_X^2)} \theta(1-\|\alpha\|^2/a_c)
\,,
\end{align}
with the following identification of the cutoff:
\begin{align}
L_c = \frac{1}{2}a_c
\,.
\end{align}

\subsection{Classical conservation laws}
\label{sec:0PM}
\begin{figure}[t]
\centering
\includegraphics[width=0.5\linewidth]{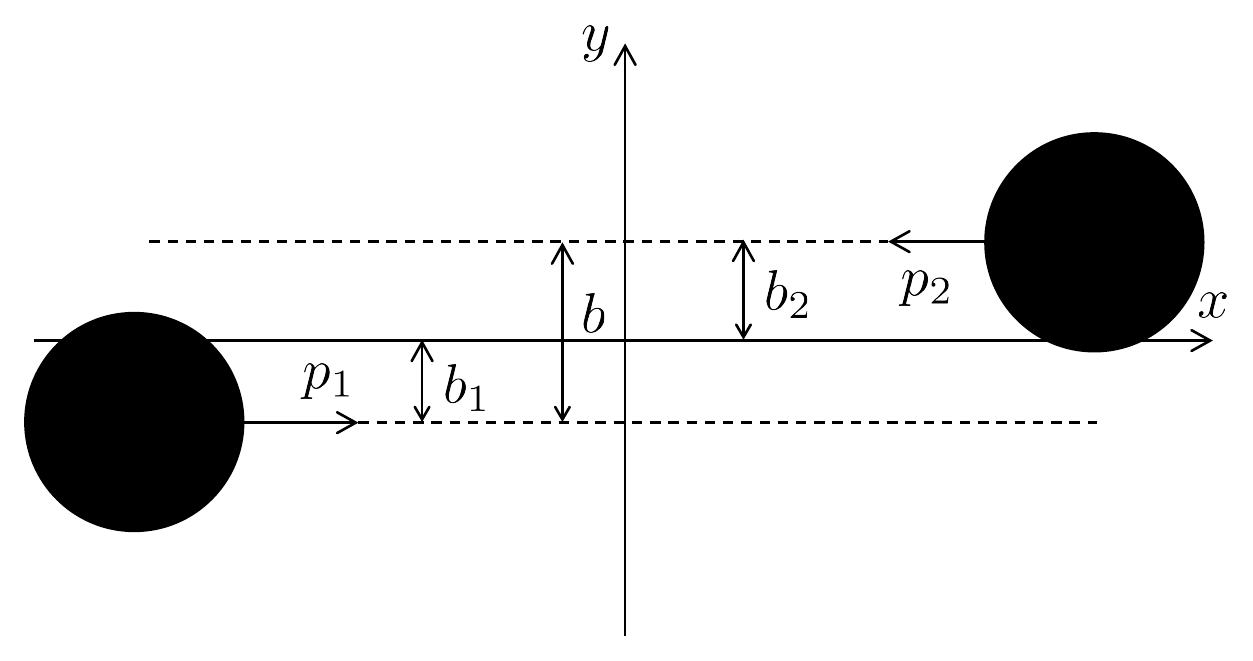}
\caption{Collision of two BHs.}
\label{fig:collision}
\end{figure}

We are now ready to compute classical observables using the KMOC formalism~\cite{Kosower:2018adc} when gravitational radiation is neglected.  Consider a problem of two incoming Schwarzschild black hole with classical four-momenta $p_1$ and $p_2$ separated by an impact parameter $b$, see Fig.~\ref{fig:collision}. When the impact parameter is smaller than the critical value $b_c=L_c/P\sim GE$, the two will collide and form a Kerr black hole. Without emission, the problem is trivial in classical physics as the four-momentum and the spin of the final state is simply determined by conservation laws. In other words, the final state is determined by the merger process in a universal way without any reference to the gravitational coupling. Nonetheless, it would be instructive to explicitly see how the classical conservation laws arise from the scattering amplitudes and their classical limit.

To simplify the calculations, in the following, we use the centre-of-mass frame and set the spatial coordinates as displayed in Fig.~\ref{fig:collision}, that is
\begin{align}
p_1^{\mu}&=(E_1, P, 0, 0) \,,  \quad
p_2^{\mu}=(E_2,-P,0, 0) \,,  \nn
b_1^{\mu}&=\left(0, 0, -\frac{E_2}{E}b, 0 \right)\,, \quad
b_2^{\mu}=\left(0,0, \frac{E_1}{E}b, 0 \right) \,, \label{kinematics} \\
b^{\mu}&=b_1^{\mu}-b_2^{\mu}=(0, 0, -b, 0) 
\,, \nonumber
\end{align}
so that the final BH forms at the origin of the centre-of-mass frame.
This specific choice of kinematics does not lose generality because this kinematics can be achieved by simply setting a coordinate system. One can perform a coordinate transformation to obtain a result for general kinematics. The classical conservation laws are
\begin{align}
    p_1^{\mu}+p_2^{\mu}=p_f^{\mu}\,, \quad L_1^{\mu\nu}+L_2^{\mu\nu}=S_f^{\mu\nu}
    \,,
\end{align}
where $L_i^{\mu\nu}=b_i^{\mu}p_i^{\nu}-b_i^{\nu}p_i^{\mu}$ is the orbital angular momentum of particle $i$ and $p_f^{\mu}$ and $S_f^{\mu\nu}$ are the four-momentum and the spin tensor of the final BH. The latter is related to the spin vector via $S_f^{\mu}=\varepsilon^{\mu}{}_{\nu\rho\sigma}u_f^{\nu}S^{\rho\sigma}_f/2=(0,0,0,bP)$ with $u_f=(1,0,0,0)$ being the four-velocity.

In the KMOC formalism, the initial state is given by two wavepackets separated by the impact parameters $b=b_1-b_2$,
\begin{align}
    \ket{\Psi}:=\int d\Phi(p_1,p_2)\phi_1(p_1)\phi_2(p_2)e^{i(b_1\cdot p_1+b_2 \cdot p_2)}\ket{p_1; p_2}
    \,,
\end{align}
with the wavefunctions $\phi_i$ being normalized to unity
\begin{align}
    \int \dd \Phi(p_i)|\phi_i(p_i)|^2 = 1
    \,.
\end{align}
The initial wavepackets should satisfy the so-called Goldilocks condition to have a well-defined classical limit:
\begin{align}
    \ell_c \ll \ell_w \ll \ell_s
    \,,
\end{align}
where $\ell_c, \ell_w,$ and $\ell_s$ are the Compton wavelength, the intrinsic spread of the wavepackets, and the separation length of two wavepackets, respectively. We are interested in the BH collisions, meaning that the separation length is of the order of the BH horizon size,
\begin{align}
    \ell_s \sim |b| \sim GE
    \,.
\end{align}
For any astrophysical black hole, the Schwarzschild radius is much larger than the Compton wavelength and thus the condition is trivially satisfied. 

The observables of our interest here are the momentum and spin of the final BH
\begin{align}
p_f^{\mu} &=\bra{\Psi}S^{\dagger}\mathbb{P}_X^{\mu}S\ket{\Psi}
\,, \label{pf} \\
S_f^{\mu} &=\bra{\Psi}S^{\dagger}\mathbb{S}_X^{\mu}S\ket{\Psi}
\,, \label{Sf}
\end{align}
where the spin operator is given in \eqref{spin_op} and the momentum operator is
\begin{align}
\mathbb{P}_X^{\mu} &:=\int \dd \Phi(X) p^{\mu}_X a^{\dagger}_X(p_X) a_X(p_X)
\,.
\end{align}
The completeness relation is given by
\begin{align}
    \mathbbm{1}&= \sum_{n=0}^{\infty}  \int_{p_1,p_2,k^n} \!\!\!  \ket{p_1;p_2; k^n}\bra{p_1;p_2;k^n} + \sum_{n=0}^{\infty}  \int_{X,k^n} \!\!\! \ket{p_X,\alpha; k^n}\bra{p_X,\tilde{\alpha};k^n} 
    + \text{($N \geq 3$ BH states)}
    \,,
    \label{completeness}
\end{align}
with the shorthand for the integrals
\begin{align}
\int_{p_1,p_2} := \int \dd \Phi(p_1) \Phi(p_2)\,, \quad 
\int_X := \int  \frac{\dd^2 \alpha \dd^2 \tilde{\alpha} }{\pi^2} \dd m_X^2 \dd \Phi(X) \rho_{\alpha}(m_X^2)
\,.
\end{align}
The $k^n$ stands for the $n$-graviton state with $\int_{k^n}$ being its phase-space integral, which we have ignored in this section. Since we are only inserting operators $O_X$, $N\geq 2$ BH states are irrelevant to our discussion. By inserting the completeness relation into \eqref{pf} and \eqref{Sf}, we obtain
\begin{align}
O_f^{\mu}&=\int_{\substack{p_1,p_2, \\ p_1',p_2',X}}  \phi_1^*(p_1')\phi_1(p_1) \phi_2^* (p_2')\phi_2(p_2)  e^{-i(b_1\cdot q_1+b_2 \cdot q_2)}  O_X^{\mu} \bra{p_1';p_2'}T^{\dagger}\ket{p_X,\alpha}\bra{p_X,\tilde{\alpha}} T \ket{p_1 ;p_2}
\,,
\end{align}
where $O_f^{\mu}=(p_f^{\mu}, S_f^{\mu})$, $O_X^{\mu}=(p_X^{\mu}, S_X^{\mu})$, and $q_i=p_i'-p_i~(i=1,2)$ are the momentum mismatches which scale as $q_i =\mathcal{O}(\hbar)$ in the classical limit. Assuming the wavefunctions are sharply peaked at the classical values of the four-momenta, the classical limit of the observables are given by~\cite{Kosower:2018adc}
\begin{align}
O_f^{\mu} &=\Dbraket{ \int_X  \prod_{i=1,2} \hat{\dd}^4 q_i \hat{\delta}(2p_i \cdot q_i) e^{-ib_i \cdot q_i}   O_X^{\mu} \bra{p_1+q_1; p_2+q_2}T^{\dagger}\ket{p_X,\alpha}\bra{p_X, \alpha}T \ket{p_1; p_2} }
\nn
&=\Dbraket{ \int_X   \mathcal{I}_3\, O_X^{\mu} \bra{p_X, \alpha}T \ket{p_1; p_2} }
\nn
&=
{\Dbraket{ \int_X   \mathcal{I}_3\, O_X^{\mu} \Amp(p_X, \alpha|p_1; p_2) \hat{\delta}^{(4)}(p_{12}-p_X) }}
\label{conservative_observables}
\end{align}
where the double-angular brackets are understood as the classical limit of the expressions inside and
\begin{align}
\mathcal{I}_3(p_1,p_2, p_X) := \int  \prod_{i=1,2} \hat{\dd}^4 q_i \hat{\delta}(2p_i \cdot q_i) e^{-ib_i \cdot q_i}   \bra{p_1+q_1; p_2+q_2}T^{\dagger}\ket{p_X,\alpha}
\label{q-integral}
\end{align}
is a Fourier transform of the three-point amplitude.

The momentum $p_X$ is fixed to be $p_X=p_1+p_2$ in \eqref{conservative_observables} because of the delta function coming from $\bra{p_X, \alpha}T \ket{p_1; p_2} $. Hence, we only need to evaluate the integral \eqref{q-integral} under $p_X=p_{12}:=p_1+p_2$,
\begin{align}
&\quad \mathcal{I}_3(p_1,p_2, p_{12}) 
\nn
&= \int  \prod_{i=1,2} \hat{\dd}^4 q_i \hat{\delta}(2p_i \cdot q_i) e^{-ib_i \cdot q_i}   \bra{p_1+q_1; p_2+q_2}T^{\dagger}\ket{p_{12},\alpha}
\nn
&=\int  \prod_{i=1,2} \hat{\dd}^4 q_i \hat{\delta}(2p_i \cdot q_i) e^{-ib_i \cdot q_i}   \hat{\delta}^{(4)}(q_1+q_2) \Amp^*(p_{12}, \tilde{\alpha} | p_1+q_1; p_2+q_2)
\nn
&=\int \hat{\dd}^4 q \hat{\delta}(2p_1 \cdot q)\hat{\delta}(2p_2\cdot q) e^{-i b\cdot q}  \Amp^*(p_{12}, \tilde{\alpha} | p_1+q; p_2-q)
\,,
\label{q-integral2}
\end{align}
where one of the $q$-integral is trivial thanks to the delta function and we have relabelled the integrated variable as $q_1 \to q$.

It is convenient for the classical limit to reintroduce the $\hbar$ dependence with the overall $\hbar^{-1/2}$ scaling of the three-point amplitude, i.e.,~$g_{\ell}\sim \hbar^{-1/2}$. This scaling is the same as the electromagnetic and gravitational couplings $e\sim \hbar^{-1/2}, \kappa \sim \hbar^{-1/2}$ and indeed gives a correct scaling to reproduce the classical physics as we will see now.
Let us first look at the $\mathcal{O}(\hbar^{-1})$ piece of the exponent of the three-point amplitude \eqref{3pt_alpha}:
\begin{align}
-\frac{1}{2\hbar}\|\alpha \|^2 +z(p_1+\hbar q,p_2-\hbar q) = -\frac{1}{2\hbar}\left[ \tilde{\alpha}_1(\alpha^1 - \tilde{\alpha}_1) + \tilde{\alpha}_2(\alpha^2 + \tilde{\alpha}_2)\right] + \mathcal{O}(\hbar^0)
\,.
\label{eq:cancel}
\end{align}
In the classical limit $\hbar \to 0$, the three-point amplitude is non-zero only when the real part of the $\mathcal{O}(\hbar^{-1})$ piece vanishes. Therefore, the $\mathcal{O}(\hbar^{-1})$ piece can be replaced with the delta function in the classical limit,\footnote{Precisely speaking, there is an imaginary part in the exponent at $\mathcal{O}(\hbar^{-1})$. However, the imaginary part vanishes with the support of the delta function and thus can be dropped safely.}
\begin{align}
\Amp = \frac{m_X}{\hbar^{1/2}} e^{-\frac{1}{2}\|\alpha \|^2 +z}  &= \hbar^{1/2} \pi E \delta_{\hbar} (\ima \alpha^1 )\delta_{\hbar} (\re \alpha^2) e^{\mathcal{O}(\hbar^0)}
\,,
\label{3pt_delta}
\end{align}
where $E=E_1+E_2=m_X$ is the centre-of-mass energy and 
\begin{align}
\delta_{\hbar}(x) = \frac{1}{\sqrt{\pi \hbar}} e^{-x^2/\hbar} \to \delta(x) \quad {\rm as}\quad \hbar \to 0
\,.
\end{align}
The appearance of the delta function is crucial to interpret \eqref{3pt_alpha} as the three-point amplitude of the classical BHs. The spin vector with the support of the delta functions \eqref{3pt_delta} and our choice of the kinematics \eqref{kinematics} is
\begin{align}
S^{\mu}_X=\left( 0, 0, (\re \alpha^1)(\ima \alpha^2), \frac{1}{2}[(\ima \alpha^2)^2 - (\re \alpha^1)^2] \right)
\,,
\label{comSX}
\end{align}
meaning that the spin is transverse to the initial momenta as it should be because the initial angular momentum must be on that plane. In other words, the so-called super-classical piece of the exponent is crucial to give rise to the classical angular momentum conservation. Yet, the spin is not completely fixed because the amplitude itself has no information about how two particles are separated. The Fourier transformation \eqref{q-integral2}, i.e., moving to position space is needed to incorporate it. Including the $\mathcal{O}(\hbar^0)$ piece, the three-point amplitude in the classical limit is 
\begin{align}
&\quad \Amp(p_{12}, \hbar^{-1/2}\tilde{\alpha} | p_1+\hbar q; p_2-\hbar q) 
\nn
& = \hbar^{1/2} \pi E \delta_{\hbar}(\ima \alpha^1 )\delta_{\hbar} (\re \alpha^2)  \exp\left[ q^x \frac{\|\alpha\|^2}{2P} - i q^y \frac{S^z_X}{P} + iq^z \frac{S^y_X}{P} \right]  
\,.
\end{align}
The integral \eqref{q-integral2} can therefore be easily performed to give
\begin{align}
\mathcal{I}_3(p_1,p_2, p_{12}) =\frac{\hbar^{5/2}}{4} \pi P \delta_{\hbar} (\ima \alpha^1 )\delta_{\hbar} (\re \alpha^2)   \delta(S^z_X - b P) \delta(S^y_X)
\end{align}
which now completely determines the spin of the $X$ state to be consistent with classical physics.

In fact, the four delta functions for $\alpha$ together with the delta functions enforcing the momentum conservation from $\bra{p_X;\alpha}T\ket{p_1;p_2}$ are sufficient to determine the classical observables~\eqref{conservative_observables}.
The $p_X$ and $m_X^2$ integrals in \eqref{conservative_observables} are straightforward thanks to the delta functions, giving
\begin{align}
O^{\mu}_f &= \Dbraket{
\hbar \pi \int \dd^2 \alpha \dd^2 \tilde{\alpha} \, \|\alpha \|^2 O_X^{\mu}|_{p_X=p_1+p_2} [ \delta_{\hbar} (\ima \alpha^1 )\delta_{\hbar} (\re \alpha^2)]^2 \delta(S^z_X - b P) \delta(S^y_X)
}
\nn
&=\frac{1}{2} \int \dd^2 \alpha \dd^2 \tilde{\alpha} \, \|\alpha \|^2 O_X^{\mu}|_{p_X=p_1+p_2} \delta (\ima \alpha^1 )\delta (\re \alpha^2) \delta(S^z_X - b P) \delta(S^y_X)
\,,
\label{eq:Of}
\end{align}
where
\begin{align}\label{eq:Squ-Gau}
[\delta_{\hbar}(x)]^2 = \frac{1}{\sqrt{\pi \hbar}} \delta_{\hbar}(\sqrt{2} x) 
\end{align}
is used before taking the $\hbar \to 0$ limit. Noting that $S_X^{\mu}$ is quadratic in $\alpha$ so there are two roots of the argument of the delta functions, the $\alpha$-integrals give the factor $2$. As a result, we obtain
\begin{align}
p_f^{\mu}=(E,0,0,0)
\,, \quad
S^{\mu}_f=(0,0,0, bP)
\,,
\end{align}
confirming the classical momentum and angular momentum conservation laws from the on-shell amplitudes.

\section{Merger amplitudes and waveforms}
\label{sec:radiation}
The three-point amplitude presented in \eqref{3pt_alpha} correctly describes a non-radiative merger process as it correctly reproduces conservation laws expected by classical physics. But how about the emission of radiation during a merger? It is natural to expect  ${\rm BH}+{\rm BH} \to {\rm BH} + {\rm gravitons}$ may be computed by this three-point and gravitational interactions.\footnote{Contact parts of higher-point amplitudes cannot be fixed by lower-point amplitudes. However, they may be irrelevant to classical observables because there is no $\hbar^{-1}$ enhancement coming from propagators unless the coupling has a stronger singularity in $\hbar \to 0$.} As usual, we can count the process by the number of gravitational coupling. Note that the radiation should also contribute to the absorption cross-section, implying that \eqref{abs_cross-section} should be replaced with
\begin{align}
\sigma^{\rm abs}_{\ell} =\sigma^{2{\rm BHs}\to {\rm BH}}_{\ell} + \sum_{n=1} \sigma^{2{\rm BHs} \to {\rm BH}+\text{$n$-gravitons}}_{\ell}
\,.
\label{abs_cross-section2}
\end{align}
Nevertheless, we may continue to use the same three-point amplitude and determine the spectral density by matching to the conditions \eqref{cross-sections} and \eqref{abs_cross-section2}. We could perturbatively solve \eqref{cross-sections} and \eqref{abs_cross-section2} for the spectral density order by order in gravitational coupling.\footnote{To compute observables in astrophysical situations, we would need a certain resummation in $G$ not only because gravity is strong around the BH but also because classical physics should correspond to the eikonal regime. Alternatively, one can employ the self-force expansion in which radiative processes should be suppressed in a large mass ratio of BHs. Then, the second term in \eqref{abs_cross-section2} should be negligible at the leading order in the self-force. See Sec.~\ref{sec:BHspacetime} for the extreme mass-ratio case.} 
Therefore, in principle, we can compute scattering amplitudes for ${\rm BH}+{\rm BH} \to {\rm BH} + {\rm gravitons}$ and associated observables based on the KMOC formalism at a certain order in the gravitational coupling. In this section, we discuss how classical observables are related to those amplitudes and then compute the waveform at $\mathcal{O}(G)$ as a first application.
\subsection{Radiative observables from on-shell amplitudes}
\label{sec:radiation_ob}
Let us begin by discussing observables of our interest and their general properties without specifying the concrete form of scattering amplitudes.
In addition to the momentum operator \eqref{pf} and the spin operator \eqref{Sf}, we are interested in the momentum radiated and waveform operators~\cite{Kosower:2018adc,Cristofoli:2021vyo}
\begin{align}
    R^{\mu} &=\bra{\Psi}S^{\dagger}\mathbb{K}^{\mu}S\ket{\Psi}\,, \qquad \mathbb{K}^{\mu} := \sum_{\sigma} \int \dd \Phi(k) k^{\mu} a_{\sigma}^{\dagger}(k) a_{\sigma}(k)
    \,, \label{Rdef}\\
    h_{\mu\nu}&=\sum_{\sigma}\kappa \int \dd \Phi(k) \varepsilon^{-\sigma}_{\mu\nu} e^{-ik\cdot x}\bra{\Psi}S^{\dagger}a_{\sigma}(k) S \ket{\Psi}
    +{\rm c.c.},
    \label{hdef}
\end{align}
where $a_{\sigma}^{\dagger}(k)$ and $a_{\sigma}(k)$ are the creation and annihilation operators of a graviton with the polarisation $\sigma=\pm$, and $\varepsilon_{\mu\nu}^{\sigma}$ is the polarisation tensor. When using the completeness relation eq.(\ref{completeness}), we may ignore the processes with increasing numbers of BHs in the classical limit as a BH cannot split into smaller BHs classically. As for the momentum \eqref{pf} and the spin \eqref{Sf} of the final BH, they are defined by the creation and annihilation operators of the $X$ state so the first term of \eqref{completeness} does not contribute. While it can in principle contribute to the momentum radiated \eqref{Rdef} and the waveform \eqref{hdef}, this contribution should vanish in the classical limit because the initial state $\ket{\Psi}$ is supposed to describe localised two particles rather than plane waves. The state $\ket{\Psi}$ has angular momentum, and when it is below the critical value $L_c$, the final state $\ket{\rm out}=S\ket{\Psi}$ should only overlap with the single final BH state in the classical limit.

Inserting the completeness relation \eqref{completeness} and omitting the first term as well as $N\geq 3$ BH states, the final BH momentum and the momentum radiated in the classical limit are given by
\begin{align}
    p_f^{\mu} &=\Dbraket{ \sum_{n=0}^{\infty} \int_{X,k^n}  p^{\mu}_X \bra{\Psi}S^{\dagger}\ket{p_X,\alpha; k^n }\bra{p_X,\tilde{\alpha}; k^n }S\ket{\Psi} }
    \,, \\
    R^{\mu} &=\Dbraket{ \sum_{n=1}^{\infty} \int_{X,k^n} k^{\mu}_n \bra{\Psi}S^{\dagger}\ket{p_X,\alpha; k^n }\bra{p_X,\tilde{\alpha}; k^n }S\ket{\Psi}}
    \,,
\label{RwithX}
\end{align}
where $k_n^{\mu}$ is the sum of momenta of radiation states. Let us use the Lorentz invariance and unitarity of S-matrix. The Lorentz invariance implies that the S-matrix is factorised into the delta function ensuring the (microscopic) momentum conservation. Hence, the final BH momentum becomes
\begin{align}
    p_f^{\mu} &=\Dbraket{ \sum_{n=0}^{\infty} \int_{X,k^n}  (p_{12}-k^{\mu}_n) \bra{\Psi}S^{\dagger}\ket{p_X,\alpha; k^n }\bra{p_X,\tilde{\alpha}; k^n }S\ket{\Psi} }
\,.
\end{align}
Here, taken together with the phase-space integral of the initial state with sharply-peaked wavefunctions, $p_1^{\mu}$ and $p_2^{\mu}$ are regarded as the classical momenta of the initial BHs. So, $p_{12}^{\mu}=p_1^{\mu}+p_2^{\mu}$ is independent of the integration variables and it can be moved outside of the integrations of $\ket{\Psi}$ and $p_X$. We again use the completeness relation \eqref{completeness} and obtain
\begin{align}
    p_f^{\mu}&=p_{12}^{\mu} \bra{\Psi}S^{\dagger}S\ket{\Psi}
    -\Dbraket{ \sum_{n=1}^{\infty} \int_{X,k^n} k^{\mu}_n \bra{\Psi}S^{\dagger}\ket{p_X,\alpha; k^n }\bra{p_X,\tilde{\alpha}; k^n }S\ket{\Psi}}
    \nn
    &=p_{12}^{\mu} - R^{\mu}
\end{align}
where unitarity $S^{\dagger}S=\mathbbm{1}$ is used. Therefore, we have confirmed that the classical momentum conservation $p_1+p_2=p_f+R$ holds if the final state does not overlap with the $N\geq 2$ BH states. The final BH loses the momentum $p_f^{\mu}$ and the mass $m_f^2=p_f^2$ by radiation emissions.\footnote{
The angular momentum conservation is less trivial in the KMOC formalism. As we have seen in the three-point amplitude, it would require figuring out the precise form of the scattering amplitude. }

The classical waveform, on the other hand, is given by~\cite{Cristofoli:2021vyo}
\begin{align}
h_{\mu\nu} = \sum_{\sigma} \kappa \int \dd \Phi(k) \varepsilon_{\mu\nu}^{-\sigma} e^{-ik\cdot x} i W^{\sigma}(p_i, k, b_i)
    +{\rm c.c.}
    \label{hmunu}
\end{align}
where
\begin{align}
    iW^{\sigma} &:=  \Dbraket{ \int \prod_{i=1,2} \hat{\dd}^4 q_i \hat{\delta}(2p_i \cdot q_i) e^{-ib_i \cdot q_i}  \bra{p_1+q_1; p_2+q_2}S^{\dagger}a_{\sigma}(k) S \ket{p_1; p_2} }
\nn
&\,= \Dbraket{ \sum_{n=0} \int_{X,k^n} \mathcal{I}_{3+n} \bra{p_X,\tilde{\alpha};k^{\sigma};k^n} T \ket{p_1; p_2} }
    \,,
\label{spectralwave}
\end{align}
is the spectral waveform with
\begin{align}
\mathcal{I}_{3+n}:= \int \prod_{i=1,2} \hat{\dd}^4 q_i \hat{\delta}(2p_i \cdot q_i) e^{-ib_i \cdot q_i}  \bra{p_1+q_1; p_2+q_2}T^{\dagger}\ket{p_X,\alpha;k^n}
\,.
\end{align}
As in the scattering case~\cite{Kosower:2018adc,Cristofoli:2021vyo}, we expect that $n=0$ gives the waveform with the radiation reaction neglected and $n\geq 1$ terms are the corrections arising from the radiation reaction.

Let us discuss the relation between the waveform and the momentum radiated. In classical physics, the momentum radiated is computed by the waveform squared while it is not manifest in \eqref{RwithX} and \eqref{spectralwave}. It would be because, in quantum theory, we can use unitarity $SS^{\dagger}=\mathbbm{1}$ and the completeness relation \eqref{completeness} for rewriting the observable in a different form; for example
\begin{align}
    R^{\mu} &= \sum_{\sigma} \int_k k^{\mu}\bra{\Psi}S^{\dagger}a^{\dagger}_{\sigma}(k) a_{\sigma}(k) S\ket{\Psi}
    \nn
    &= \sum_{\sigma}\int_k k^{\mu}\bra{\Psi}S^{\dagger}a^{\dagger}_{\sigma}(k) SS^{\dagger} a_{\sigma}(k) S\ket{\Psi}
    \nn
     &=\sum_{\sigma} \int_{k,p_1,p_2} k^{\mu}\bra{\Psi}S^{\dagger}a^{\dagger}_{\sigma}(k) S\ket{p_1; p_2}\bra{p_1; p_2} S^{\dagger} a_\sigma(k) S\ket{\Psi} +\cdots 
     \label{R_transform}
\end{align}
where the completeness relation is inserted between $S$ and $S^{\dagger}$ to obtain the last line and $\cdots$ are the contributions from other than two BH states. If the ellipsis vanishes in the classical limit, we can indeed show that the momentum radiated is given by the square of the spectral waveform
\begin{align}
R^{\mu} = \sum_{\sigma} \int_k k^{\mu} |W^{\sigma} |^2
\,,
\label{RwithW}
\end{align}
in the classical limit. Another way to reach the form \eqref{RwithW} is to postulate that the final state $S\ket{\Psi}$ is given by a coherent state of gravitons in the classical limit~\cite{Cristofoli:2021vyo}. The spectral waveform is nothing but the eigenvalue of the coherent state and it immediately leads to \eqref{RwithW}. While the general proof of the equivalence of two representations \eqref{RwithX} and \eqref{RwithW} is beyond the scope of the present paper, we will argue that they can give the same answer by using the explicit form of scattering amplitudes.

\subsection{Four-point amplitudes at leading order}
As a first application of the on-shell approach to BH mergers, we compute the leading-order waveform given by the tree-level four-point amplitude with two incoming spin-0 particles and outgoing spin-$\ell$ particle and graviton. We write this amplitude as
\begin{align}
\Amp(\bm{X}^{\ell}; k^{\pm}|p_1; p_2) = g_{\ell} \frac{\kappa}{2}\tilde{\Amp}_4^{\pm}
\end{align}
where $ \tilde{\Amp}_4^{\pm}$ denotes the kinematic dependence of the four-point amplitude. Combined with the overall $\kappa$ of \eqref{hmunu}, the waveform will be linear in $G$ and hence 1PM.
The four-point is obtained by the tree-level gluing of three-point amplitudes. The three-point amplitude of three massive particles has been already explained in Sec.~\ref{sec:3pt_ell}. We assume that the coupling to a graviton is given by the minimal coupling~\cite{Arkani-Hamed:2017jhn}
\begin{align}\label{eq: 3point}
\begin{tikzpicture}[baseline=-2]
\begin{feynhand}
\propag (0.6,0) node [right] {$p_i'{}^{\ell}$} -- (0, 0);
\propag (-0.6, -0.6) node [left] {$p_i^{\ell}$} -- (0, 0);
\propag [boson] (-0.6,0.6) node [left] {$k^{+2}$} -- (0,0)  ;
\end{feynhand}
\end{tikzpicture}
=\frac{\kappa}{2} m_i^2 x_i^2 (\epsilon_{\alpha\beta})^{2\ell}
\,, \quad
\begin{tikzpicture}[baseline=-2]
\begin{feynhand}
\propag (0.6,0) node [right] {$p_i'{}^{\ell}$} -- (0, 0);
\propag (-0.6, -0.6) node [left] {$p_i^{\ell}$} -- (0, 0);
\propag [boson] (-0.6,0.6) node [left] {$k^{-2}$} -- (0,0)  ;
\end{feynhand}
\end{tikzpicture}
=\frac{\kappa}{2}\frac{m_i^2}{x_i^2} (\epsilon_{\dot{\alpha}\dot{\beta}})^{2\ell}
\,,
\end{align}
where the $x$-factor is defined to satisfy
\begin{align}
x_i \ket{k} = \frac{p_i|k]}{m_i}\,, \quad
\frac{|k]}{x} = \frac{p_i \ket{k}}{m_i}
\,.
\label{x_propto}
\end{align}
The $x$-factor cannot be expressed in a local way, and this non-locality is crucial to find the four-point amplitude to have a consistent factorisation. 

In the following, we take the positive helicity to explain how to obtain the four-point amplitude. Let's first compute the $t$-channel glueing:
\begin{align}
\begin{tikzpicture}[baseline=0]
\begin{feynhand}
\vertex (a) at (0,0) {$\otimes$} node [right] {$~p$}; 
\propag (0.8, 1.2) node [right] {$1$} --(0,0.8) -- (a) ;
\propag (a) -- (0,-0.8) -- (0.8,-1.2) node [right] {$2$};
\propag (-0.8,-1.2) node [left] {$X^{\ell}$} -- (0, -0.8);
\propag [boson] (-0.8,1.2) node [left] {$k^{+2}$} -- (0,0.8)  ;
\end{feynhand}
\end{tikzpicture}
\qquad 
 \begin{aligned} 
R_t &=x_{1}^2m_1^2 \bra{\bm X}p p_2 \ket{\bm X}^{\ell}
\\
&=x_{1}^2 m_1^2 \left(\bra{\bm X}p_1  p_2 \ket{\bm X} - \braket{{\bm X}k}[k|p_2\ket{\bm X} \right)^{\ell}
\,.
\label{t-channel1}
\end{aligned}
\end{align}
The non-local $x$-factor has a pole and it must be correctly interpreted as a pole of another channel~\cite{Arkani-Hamed:2017jhn}. To do so, we expand the tensor product of \eqref{t-channel1} and use the relations \eqref{x_propto} to separate the $x$-dependent and independent pieces
\begin{align}
R_t&=x_{1}^2 m_1^2 \Biggl[ \bra{\bm X}p_1  p_2 \ket{\bm X} ^{\ell} - \ell \bra{\bm X}p_1  p_2 \ket{\bm X}^{\ell-1}  \braket{{\bm X}k}[k|p_2\ket{\bm X} 
\nn
&\qquad \qquad + \sum_{n=2}^{\ell} (-1)^n \binom{\ell}{n} \bra{\bm X}p_1 p_2 \ket{\bm X}^{\ell -n} (\braket{{\bm X}k}[k|p_2\ket{\bm X} )^n
\Biggl]
\nn
&=x_{1}^2 m_1^2 \bra{\bm X}p_1  p_2 \ket{\bm X} ^{\ell} -\ell  x_{1} m_1 \bra{\bm X}p_1  p_2 \ket{\bm X}^{\ell-1}  \bra{\bm X} p_1| k] [k|p_2\ket{\bm X}  
\nn
&+\sum_{n=2}^{\ell} (-1)^n \binom{\ell}{n} \bra{\bm X}p_1 p_2 \ket{\bm X}^{\ell -n} ( \braket{{\bm X}k} [k|p_2 \ket{\bm X} )^{n-2} (\bra{\bm X}p_1 |k][k|p_2 \ket{\bm X})^2
\,.
\end{align}
The four-point amplitude must be symmetric in $1\leftrightarrow 2$ as both are spin-0 particles. We thus interpret the $x$-factors as having the $u$-channel pole, that is
\begin{align}
m_1 x_1 &= m_1 x_1 \frac{[k|p_2\ket{k} }{[k|p_2\ket{k}} = -\frac{[k|p_1p_2|k] }{2k \cdot p_2} = \frac{[k|p_1p_2 |k]}{u-m_2^2}
\,, \\
m_1^2x_1^2 &=  m_1 x_1 \frac{[k|p_1p_2|k]}{u-m_2^2} = - \frac{[k|p_1p_2 |k]}{u-m_2^2} \frac{[k|p_1p_X |k]}{s-m_X^2} = - \frac{[k|p_1p_2 |k]^2}{(u-m_2^2)(s-m_X^2)} 
\,,
\end{align}
where $[k|p_1p_X |k] = [k|p_1(p_1+p_2-k)|k] = [k|p_1p_2|k]$ is used. The Mandelstam variables are
\begin{align}
    s=(p_1+p_2)^2=(p_X+k)^2 \,, \quad t=(p_1-k)^2\,, \quad  u=(p_2-k)^2
    \,.
\end{align}
Therefore, the $t$-channel residue can be written as
\begin{align}
R_t &= -\frac{[k|p_1  p_2|k]^2 \bra{\bm X}p_1 p_2 \ket{\bm X}^{\ell}}{(u-m_2^2)(s-m_X^2)} -\frac{ \ell \bra{\bm X}p_1 p_2 \ket{\bm X}^{\ell-1} \bra{\bm X}p_1 |k][k|p_2 \ket{\bm X} [k| p_1 p_2 |k]}{u-m_2^2}
\nn
&+\sum_{n=2}^{\ell} (-1)^n \binom{\ell}{n} \bra{\bm X}p_1 p_2 \ket{\bm X}^{\ell -n} ( \braket{{\bm X}k} [k|p_2 \ket{\bm X} )^{n-2} (\bra{\bm X}p_1 |k][k|p_2 \ket{\bm X})^2
\,,
\end{align}
suggesting that the four-point amplitude is given by
\begin{align}
\tilde{\Amp}^+_4 &=
-\frac{[k|p_1  p_2|k]^2 \bra{\bm X}p_1 p_2 \ket{\bm X}^{\ell}}{(t-m_1^2)(u-m_2^2)(s-m_X^2)} - \frac{ \ell \bra{\bm X}p_1 p_2 \ket{\bm X}^{\ell-1} \bra{\bm X}p_1 |k][k|p_2 \ket{\bm X} [k| p_1 p_2 |k]}{(t-m_1^2)(u-m_2^2)}
\nn
&+\left[ \sum_{n=2}^{\ell} (-1)^n \binom{\ell}{n} \bra{\bm X}p_1 p_2 \ket{\bm X}^{\ell -n} ( \braket{{\bm X}k} [k|p_2 \ket{\bm X} )^{n-2} (\bra{\bm X}p_1 |k][k|p_2 \ket{\bm X})^2\right] \frac{1}{t-m_1^2}
\nn
&+ \Bigl[ \quad (1\leftrightarrow 2) \quad \Bigl] \frac{1}{u-m_2^2}
\,.
\label{4ptA}
\end{align}
This amplitude matches the $u$-channel residue because of the manifest $1\leftrightarrow 2$ symmetry. The $s$-channel residue of \eqref{4ptA} is
\begin{align}
    R_s&=-\frac{[k|p_1  p_2|k]^2 \bra{\bm X}p_1 p_2 \ket{\bm X}^{\ell}}{(t-m_1^2)(u-m_2^2)}
    \nn
    &=m_X^2x_X^2  \bra{\bm X}p_1 p_2 \ket{\bm X}^{\ell}
\end{align}
which exactly matches the $s$-channel with the minimal coupling of the spin-$\ell$ particles and the graviton. Therefore, the amplitude \eqref{4ptA} correctly has the factorisation properties of all channels with minimal coupling to the graviton. 
The four-point amplitude with a negative helicity graviton is computed similarly and it is simply given by \eqref{4ptA} with the replacement $\ket{\,\cdot\,} \leftrightarrow |\,\cdot\,]$.
One might be curious what happens if the graviton is non-minimally coupled to the spin-$\ell$ field. It turns out such coupling appearing in the $s$-channel will not affect $t,u$-channel residues and hence is invisible from our $t,u$-channel on-shell construction. To see this consider modifying eq.\eqref{eq: 3point} with terms of the form $x_i^2 \epsilon^{2\ell{-}a}(x_i\lambda_k\lambda_k)^{a}$ for $a\in even$ where $\lambda_k$ is the massless spinor. For $a\neq 0$ these are non-minimal couplings. For $a\geq 2$ we can simply rewrite
\begin{align}
x_i^2 (x_i\lambda_k\lambda_k)^{a}=\left(\frac{p_i|k]}{m_i}\frac{p_i|k]}{m_i}\right)^{2}\left(\frac{p_i|k]}{m_i}\lambda_k\right)^{a{-}2}\,.
\end{align}
Since all $x_i$s are removed, such terms do not introduce $t,u$-channel poles.

Having computed the 4-point amplitudes, we multiply $\tilde{\alpha}_I$ and use the coupling constant \eqref{gell} to find
\begin{align}
\frac{2}{\kappa}\frac{\ell !}{\sqrt{(2\ell)!}} \tilde{\alpha}^{2\ell} \Amp^{\pm}
&= 
-\frac{(A^{\pm})^2 (z^{\pm})^{\ell}}{(t-m_1^2)(u-m_2^2)(s-m_X^2)} - \frac{ \ell A^{\pm} v^{\pm} (z^{\pm})^{\ell-1}  }{(t-m_1^2)(u-m_2^2)}
\nn
&+ (v^{\pm})^2\sum_{n=2}^{\ell} (-1)^n \binom{\ell}{n}\left[  \frac{(z^{\pm})^{\ell -n} (w^{\pm}_2)^{n-2}  }{t-m_1^2} +\frac{(z^{\pm})^{\ell -n} (w^{\pm}_1)^{n-2} }{u-m_2^2} \right]
\,, 
\end{align}
where
\begin{alignat}{2}
A^+&:=[k|p_1p_2|k]\,, \qquad 
&A^-&:=\bra{k}p_1p_2\ket{k}
\,, \\
z^+&:=\frac{\tilde{\alpha}_I\bra{X^{I}}p_1 p_2 \ket{X^{J}} \tilde{\alpha}_J}{ m_X \lambda^{1/2}(m_1^2,m_2^2,m_X^2)}
\,, \qquad 
 &z^-&:=\frac{\tilde{\alpha}_I[X^{I}|p_1 p_2 |X^{J}] \tilde{\alpha}_J}{ m_X \lambda^{1/2}(m_1^2,m_2^2,m_X^2)}
 \,,
\\
v^+& :=  \frac{\tilde{\alpha}_I \bra{X^I }p_1 |k][k|p_2 \ket{X^J}\tilde{\alpha}_J }{m_X \lambda^{1/2}(m_1^2,m_2^2,m_X^2) }
\,,  \qquad 
&v^- &:=  \frac{\tilde{\alpha}_I [X^I|p_1 \ket{k}\bra{k}p_2 |X^J]\tilde{\alpha}_J }{m_X \lambda^{1/2}(m_1^2,m_2^2,m_X^2) }\,,
\\
w^+_i &:= \frac{  \tilde{\alpha}_I \braket{X^I k} [k|p_i \ket{ X^J}\tilde{\alpha}_J  }{ m_X \lambda^{1/2}(m_1^2,m_2^2,m_X^2) }
\,, \qquad 
&w^-_i &:= \frac{  \tilde{\alpha}_I [X^I k] \bra{k}p_i | X^J]\tilde{\alpha}_J  }{ m_X \lambda^{1/2}(m_1^2,m_2^2,m_X^2) }
\,.
\end{alignat}
Note that $z^{\pm}$ reduces to $z$ defined by \eqref{zdef} in the three-point kinematics,
\begin{align}
    z^{\pm}|_{p_X=p_{12}}=z
    \,.
\end{align}
The infinite sum \eqref{def_alpha} is easily performed to give the tree-level minimal-coupling four-point amplitude of the coherent spin state as follows:
\begin{align}
\Amp(p_X,\tilde{\alpha}; k^{\pm}|p_1 ; p_2) &=\Amp(p_{12},\tilde{\alpha}|p_1; p_2) S^{\pm}_{\alpha}
\,,
\label{4pt_alpha}
\end{align}
where
\begin{align}
S^{\pm}_{\alpha}=\frac{\kappa}{2} e^{\Delta z^{\pm}} \Biggl[ &-\frac{(A^{\pm})^2 }{(t-m_1^2)(u-m_2^2)(s-m_X^2)} -\frac{ A^{\pm} v^{\pm} }{(t-m_1^2)(u-m_2^2)}
\nn
& - \frac{(v^{\pm})^2(1-w_2^{\pm} -e^{-w_2^{\pm}}) }{(w_2^{\pm})^2(t-m_1^2)} +\frac{(v^{\pm})^2(1- w_1^{\pm} -e^{-w_1^{\pm}}) }{(w^{\pm}_1)^2(u-m_2^2)} \Biggl]
\,,
\label{Splus}
\end{align}
with
\begin{align}
    \Delta z^{\pm}:=z^{\pm}(p_1,p_2,p_{12}-k)-z(p_1,p_2) \,.
\end{align}
The four-point amplitude \eqref{4pt_alpha} is factorised into the three-point \eqref{3pt_alpha} and the factor $S^{\pm}_{\alpha}$ à la soft theorem. Note that this factorisation does not require any assumption on the size of the spin $\alpha$ and shows up even before taking the soft limit thanks to the use of the coherent spin state.

\subsection{All-order gravitational spin memory}

We use the obtained four-point \eqref{4pt_alpha} and the KMOC formalism to compute the gravitational waveform at leading order. The leading spectral waveform is given by
\begin{align}
iW^{\sigma} = \Dbraket{\int_X \mathcal{I}_3(p_1,p_2,p_{12}-k) \bra{p_X,\tilde{\alpha};k^{\sigma}}T\ket{p_1;p_2} }
\end{align}
where the Fourier integral $\mathcal{I}_3$ needs to be evaluated at the four-particle kinematics $p_X=p_{12}-k$:
\begin{align}
&\quad \mathcal{I}_3(p_1,p_2, p_{12}-k) 
\nn
&= \int  \prod_{i=1,2} \hat{\dd}^4 q_i \: \hat{\delta}(2p_i \cdot q_i) e^{-ib_i \cdot q_i}   \hat{\delta}^{(4)}(q_1+q_2+k) \Amp^*(p_{12}-k, \tilde{\alpha} | p_1 +q_1; p_2+q_2)
\nn
&= \int  \hat{\dd}^4 q \: \hat{\delta}(2p_1 \cdot q) \hat{\delta}(2p_2\cdot (q +k) )  e^{-i b \cdot q + i b_2 \cdot k}  \Amp^*(p_{12}-k, \tilde{\alpha} | p_1 +q; p_2-q-k)
\,.
\label{I3_p12-k}
\end{align}
We use the same kinematics \eqref{kinematics} and the parametrisation of the massive spinor \eqref{massive_spinor1}, giving
\begin{align}
&\quad \Amp(p_{12}-k, \tilde{\alpha} | p_1 +q; p_2-q-k)  
\nn
&=  \pi E \delta_{\hbar}(\ima \alpha^1 )\delta_{\hbar} (\re \alpha^2)  
\nn
&\times \exp\left[ \left(q^x + \frac{E_1E_2}{EP}\omega + \frac{E_1}{E}k^x \right) \frac{\|\alpha\|^2}{2P} - i \left(q^y+\frac{E_1}{E}k^y \right) \frac{S^z_X}{P} + i \left(q^z +\frac{E_1}{E}k^z \right) \frac{S^y_X}{P} \right]  
\end{align}
in the classical limit. Here, $\omega$ is the frequency of the graviton, $k^{\mu}=(\omega, k^x, k^y,k^z)$. It is straightforward to perform the $q$ integral of \eqref{I3_p12-k}, obtaining the same expression as before,
\begin{align}
\mathcal{I}_3(p_1,p_2, p_{12}-k) =\frac{1}{4} \pi P \delta_{\hbar} (\ima \alpha^1 )\delta_{\hbar} (\re \alpha^2)   \delta(S^z_X - b P) \delta(S^y_X)
\,,\label{eq:I3}
\end{align}
in our choice of the kinematics \eqref{kinematics}.

Following the same calculation in Sec.~\ref{sec:0PM} and using the factorisation \eqref{4pt_alpha}, one can recognise that the spectral waveform will be given by the factor $S^{\pm}_{\alpha}$. This is reminiscent of the well-known relation between the gravitational (spin) memory and the (sub-leading) soft graviton theorem~\cite{Strominger:2014pwa, Pasterski:2015tva} (see also~\cite{Laddha:2018rle, Laddha:2019yaj}), and indeed $S^{\pm}_{\alpha}$ gives the gravitational memory as we will see below. In the following, we elaborate on how the soft factor gives rise to the gravitational memory purely from the perspective of the on-shell scattering amplitudes.

Let us take the classical limit of the four-point \eqref{4pt_alpha}. We recall that the massive three-point becomes the delta function \eqref{3pt_delta} ensuring (a part of) angular momentum conservation laws. We can use the support of the delta function to take the classical limit of the factor $S_{\alpha}^{\pm}$. While our computation is based on the centre-of-mass coordinates \eqref{kinematics}, the amplitude is Lorentz invariant so we may recover the covariant expressions from coordinate expressions. In the classical limit, the result must be a function of classical kinematical quantities, namely the momenta $p_1,p_2,k$, the polarisation vectors $\varepsilon^{\pm}$, and the classical spin $S_X$. The spin $S_X$ is given by the form \eqref{comSX} in our kinematics thanks to the support of the factorised massive three-point $\Amp(p_{12},\tilde{\alpha}|p_1;p_2)$. The momentum and mass of the $X$ state are $p_X=p_{12}-k$ and $m_X^2=s-2k\cdot p_{12}$ by the conservation law. The polarisation vectors are
\begin{align}
    \varepsilon^+_{\mu}=\frac{\bra{\xi}\sigma_{\mu}|k]}{\sqrt{2}\braket{\xi k}}
    \,, \quad
    \varepsilon^-_{\mu}=-\frac{[\xi|\bar{\sigma}_{\mu}\ket{k}}{\sqrt{2}[\xi k]}
\end{align}
with $\xi$ being a reference null vector. The gauge invariance of the amplitude is ensured by the independence of the reference. Then, using these building blocks, we find that our expression may be covariantised into
\begin{align}
    \Delta z^{\pm} &= \pm k \cdot a_X + 2 k\cdot p_{12} |S_X| \frac{s(s-m_1^2-m_2^2)}{\lambda(s,m_1^2,m_2^2)} + \mathcal{O}(\hbar)\,,
    \\
    v^{\pm} &=\pm \frac{4\sqrt{2} \varepsilon^{\pm}_{[\mu} k_{\nu]}}{\lambda(s,m_1^2,m_2^2)}\Big[ 2im_1^2  S_X^{\mu}{}_{\rho}p_2^{\rho}p_2^{\nu}
    -2im_2^2  S_X^{\mu}{}_{\rho}p_1^{\rho}p_1^{\nu} 
    +(s-m_1^2-m_2^2)|S_X| p_1^{\mu}p_2^{\nu}
    \Big] + \mathcal{O}(\hbar) 
    \,, \\
    w_1^{\pm} &=\mp k\cdot a_X + \frac{4}{\lambda(s,m_1^2,m_2^2)}\Big[
    i (p_1\cdot p_{12}) k_{\mu}S_X^{\mu\nu}p_{1\nu}
    +2 p_1^{[\mu} p_{2}^{\nu]} p_{1\mu}k_{\nu}|S_X|\Big] + \mathcal{O}(\hbar)
    \,, \\
    w_2^{\pm} &=\pm k\cdot a_X - \frac{4}{\lambda(s,m_1^2,m_2^2)}\Big[
    i (p_2\cdot p_{12}) k_{\mu}S_X^{\mu\nu}p_{2\nu}
    +2 p_2^{[\mu} p_{1}^{\nu]} p_{2\mu}k_{\nu}|S_X|\Big] + \mathcal{O}(\hbar)
    \,,
\end{align}
on the support of the delta function where $|S_X|=\sqrt{-S_X \cdot S_X}$ and $a_X^{\mu}:=S_X^{\mu}/m_X$ is the spin-length vector. Note that $\Delta z^{\pm}, v^{\pm}, w_j^{\pm}$ are all $\mathcal{O}(\hbar^0)$. 
All in all, by writing the kinematic dependence explicitly, the classical limit of the four-point is written as
\begin{align}
 \lim_{\hbar \to 0}\Amp(p_X,\tilde{\alpha};k^{\pm}|p_1;p_2)=\lim_{\hbar \to 0}\Amp(p_{12},\tilde{\alpha}|p_1;p_2)S^{\pm}_{\alpha}(p_1,p_2,k,S_X)
 \label{4pt_cl}
\end{align}
with $S_X$ being subject to orthogonal to $p_1$ and $p_2$ but not yet fixed. The $\hbar$ scaling of $S^{\pm}_{\alpha}$ is $\hbar^{-3/2}$ where $\hbar^{-1/2}$ comes from the gravitational coupling $\kappa$ and $\hbar^{-1}$ is from the propagator.

We proceed to compute the waveform,
\begin{align}
    iW^{\pm}= \Dbraket{\int_X \mathcal{I}_3(p_1,p_2,p_X) \Amp(p_{12},\tilde{\alpha}|p_1;p_2)S^{\pm}_{\alpha}(p_1,p_2,k,S_X) \hat{\delta}^{(4)}(p_{12}-p_X-k) }
    \,.
    \label{Wpm}
\end{align}
As {shown in eq.~\eqref{eq:I3}}, the Fourier integral $\mathcal{I}_3$ in the four-point kinematics $p_{12}-p_X-k=0$ gives the same answer as the three-point kinematics $p_{12}-p_X=0$ in our centre-of-mass coordinates \eqref{kinematics}. Since the other parts of the integrand have no cancellation at the leading order of $\hbar$, we can safely ignore $k$ in the delta function to obtain
\begin{align}
iW^{\pm}= \Dbraket{\int_X \mathcal{I}_3(p_1,p_2,p_X) \Amp(p_{12},\tilde{\alpha}|p_1;p_2)S^{\pm}_{\alpha}(p_1,p_2,k,S_X) \hat{\delta}^{(4)}(p_{12}-p_X) }
    \,.
\end{align}
This equation should be compared with \eqref{conservative_observables}; the only difference is that $\mathcal{O}_X^{\mu}$ in \eqref{conservative_observables} is replaced with $S^{\pm}_{\alpha}(p_1,p_2,k,S_X)$. We recall how the classical observables $O_f^{\mu}$ were derived from the $X$-state quantity $\mathcal{O}_X^{\mu}$ through the integration \eqref{conservative_observables}. The point was that the Fourier integral $\mathcal{I}_3$ together with the delta function of three-point $\hat{\delta}^{(4)}(p_{12}-p_X) $ gives enough Dirac delta functions to algebraically perform all the $X$ integrals in the classical limit, which fixes the kinematics of $X$ in terms of the initial kinematics according to the classical conservation. Therefore, we end up with
\begin{align}
    iW^{\pm}=\lim_{\hbar \to 0} S_{\alpha}^{\pm} |_{S_X=L_1+L_2}
    \,,
    \label{W=S}
\end{align}
with $L_i^{\mu\nu}=b_i^{\mu}p_i^{\nu}-b_i^{\nu}p_i^{\mu}$ being the initial kinematics orbital angular momenta. $W^{\pm}=\mathcal{O}(\hbar^{-3/2})$ is indeed the correct $\hbar$ scaling to have classical GWs~\cite{Cristofoli:2021vyo}. Eq.~\eqref{W=S} gives an on-shell proof of the equivalence between the GW memory and the soft factor\footnote{See also \cite{Bautista:2021llr, Bautista:2019tdr} for an on-shell application on $2 \rightarrow 2$ scattering processes.}. The soft limit is included in $\hbar \to 0$ because the momentum of the graviton scales as $k=\mathcal{O}(\hbar)$. On top of that, the KMOC formalism and the coherent spin formalism specify how other variables should scale in taking the limit $\hbar \to 0$, leading to the precise relation between the classical waveform and the quantum scattering amplitudes at all orders in the spin.

It is instructive to explicitly see lower spin orders of the waveform. Up to $\mathcal{O}(\omega^1 S^2)$, we find that \eqref{W=S} can be expressed as the universal form of the {\it classical} version of the sub-subleading soft graviton operator~\cite{Cachazo:2014fwa, Pasterski:2015tva, Laddha:2019yaj, Laddha:2017ygw, Laddha:2018rle}
\begin{align}
    iW^{\pm} &=\frac{\kappa}{2}\sum_{i=1}^3\left[\frac{(\varepsilon^{\pm} \cdot p'_i)^2}{k\cdot p'_i} - i \frac{(\varepsilon^{\pm} \cdot p'_i)(\varepsilon^{\pm} \cdot J_i \cdot k)}{k\cdot p'_i} -\frac{1}{2}\frac{(\varepsilon^{\pm} \cdot J_i \cdot k)^2}{k \cdot p'_i} \right] + \mathcal{O}(\omega^2 S^3)
    \,,
    \label{soft_theorem}
\end{align}
in the all outgoing notation: the momenta are $p'{}^{\mu}_{1,2} = -p^{\mu}_{1,2}$, $p'{}^{\mu}_3=p^{\mu}_f$ while the total angular momentum of each particle is $J^{\mu\nu}_{1,2}=-L^{\mu\nu}_{1,2}, J^{\mu\nu}_3=S^{\mu\nu}_f$ with the conservation law $\sum_{i=1}^3 J^{\mu\nu}_i=0$ in our kinematics \eqref{kinematics}. This waveform can be obtained by pure classical calculations at the subleading order~\cite{Laddha:2019yaj} or by replacing quantum angular momentum operators of the tree-level gravity amplitude with the classical ones~\cite{Cachazo:2014fwa}. However, it is non-trivial why we can replace the quantum operator, especially with the orbital angular momentum because the momentum space amplitudes do not know the separation of two particles. In fact, we have seen that the four-point amplitude~\eqref{4pt_cl} itself contains the information on the final spin $S_X$ only. The observable \eqref{Wpm} requires a multiplication of the Fourier transform of the massive three-point and, after performing the phase-space integral, the final spin is fixed in terms of the initial orbital angular momenta. Then, we are allowed to interpret $S_X=S_f$ as $L_1+L_2$ and can distribute them to reach the universal form \eqref{soft_theorem}. In other words, the ``in-in'' structure of the observable $\bra{\Psi}S^{\dagger}a_{\sigma}(k) S \ket{\Psi}$ is essential to justify the replacement from the quantum operators to the classical quantities.

Let us briefly mention how the gravitational memory waveform with spin is obtained in purely classical calculation~\cite{Laddha:2019yaj} (see also \cite{Ghosh:2021bam}). It starts with separating the space region into the free region $|\vec{x}|>L$ and the interaction region $|\vec{x}|<L$ by assuming all the interactions take place in a local region $|\vec{x}|<L$. The final result must be independent of this artificial separation so the boundary contribution of $|\vec{x}|=L$ arising from one side should be cancelled with the contribution from the other side. The energy-momentum tensor in the free region is given by the sum of $T_{\mu\nu}$ of a free (spinning) particle, which is unambiguously given once multipole moments of the particle are fixed. On the other hand, $T_{\mu\nu}$ in the interaction region generically requires knowledge of the interactions. Nevertheless, the low-frequency part of the emitted GWs must be insensitive to the short-distance physics and there may be a universal way to describe $T_{\mu\nu}(|\vec{x}|<L)$ for low frequencies. In fact, \cite{Laddha:2019yaj} found a way to determine $T_{\mu\nu}(|\vec{x}|<L)$ from the conservation law at the subleading order in spin. Then, adding both $T_{\mu\nu}$ in the free and interaction regions, the spin memory waveform \eqref{soft_theorem} is obtained up to the subleading order. However, as noted in~\cite{Laddha:2019yaj}, ambiguities associated with the interaction region, such as the presence of irrelevant interactions,  $T_{\mu\nu}(|\vec{x}|<L)$ will modify the sub-subleading order. 

In our on-shell approach, there was no ambiguity to extend the memory to higher spin orders. It traces back to the fact that the gravitational multipole moments for Kerr BH are given by a minimal coupling. This fixes the four-point amplitude completely, with the memory unambiguously extended into all spin orders. In other words, the on-shell method has led to universal and systematic investigations of interactions, and together with the simplicity of the Kerr BH, it enables us to completely determine the memory associated with the merger of  Schwarzschild BHs into a Kerr BH. We, however, stress that our result is all order in the spin but linear in the gravitational coupling. The waveform is valid only when loop amplitudes, namely higher contributions are negligible. Since $G$ has negative mass dimensions, the dimensionless expansion parameter requires the presence of at least one quantity with positive mass dimensions. It is natural to expect that the frequency plays a role, and therefore, our waveform should be valid only in the small-frequency limit. In fact, if there were no UV cutoff in the frequencies, the time domain waveform would be ill-defined because of the exponential function in \eqref{Splus}.

Let us briefly discuss the radiated momentum. In Sec.~\ref{sec:radiation_ob}, we discussed that the momentum radiated can be represented in two different ways, \eqref{RwithX} and \eqref{RwithW}. Let us see the equivalence of the two at the leading order. We shift $p_1$ and $p_2$ of \eqref{4pt_cl} by of the order of $\hbar$. Since $S_{\alpha}^{\pm}$ does not have the super-classical piece, this shift does not alter the soft factor in the classical limit. Therefore, the $\hbar$ shift only affects the massive three-point,
\begin{align}
\Amp(p_X,\tilde{\alpha};k^{\pm}|p_1+q_1;p_2+q_2)=\Amp(p_{12}+q_{12},\tilde{\alpha}|p_1+q_1;p_2+q_2)S^{\pm}_{\alpha}(p_1,p_2,k,S_X)
\end{align}
where the classical limit is understood. The Fourier transform of the four-point amplitude on the support of the four-particle kinematics $p_{12}=p_X+k$ is
\begin{align}
    \mathcal{I}_4 &=
    \int \prod_{i=1,2} \hat{\dd}^4 q_i \hat{\delta}(2p_i \cdot q_i) e^{-ib_i \cdot q_i}  \bra{p_1+q_1; p_2+q_2}T^{\dagger}\ket{p_X,\alpha;k^{\pm}}
    \nn
    &=\int \prod_{i=1,2} \hat{\dd}^4 q_i \hat{\delta}(2p_i \cdot q_i) e^{-ib_i \cdot q_i}  \hat{\delta}^{(4)}(q_1+q_2)\Amp^*(p_{12},\tilde{\alpha}|p_1+q_1;p_2+q_2) (S^{\pm}_{\alpha})^*
    \nn
    &=\mathcal{I}_3(p_1,p_2,p_{12}) (S^{\pm}_{\alpha})^*
    \,,
\end{align}
and then the leading-order momentum radiated is
\begin{align}
R^{\mu} &= \sum_{\sigma} \Dbraket{ \int_k k^{\mu} \int_X \mathcal{I}_4 \,\Amp(p_X,\tilde{\alpha};k^{\sigma}|p_1;p_2) \hat{\delta}^{(4)}(p_X+k-p_{12})}
\nn
&=\sum_{\sigma}\Dbraket{ \int_k k^{\mu} \int_X \mathcal{I}_3(p_1,p_2,p_{12}) \Amp(p_X,\tilde{\alpha}|p_1;p_2) |S_{\alpha}^{\pm}|^2 \hat{\delta}^{(4)}(p_X+k-p_{12})}
\nn
&=\sum_{\sigma}\Dbraket{ \int_k k^{\mu} \int_X \mathcal{I}_3(p_1,p_2,p_{12}) \Amp(p_X,\tilde{\alpha}|p_1;p_2) |S_{\alpha}^{\sigma}|^2 \hat{\delta}^{(4)}(p_X-p_{12})}
\nn
&=\sum_{\sigma}\int_k k^{\mu} |W^{\sigma}|^2
\,.
\label{RX=RW}
\end{align}
The integrand of momentum radiated is indeed given by the waveform squared. Note that the $k$ integral is not well-defined for the same reason as the time domain waveform unless a UV cutoff is introduced. It would be desirable to calculate the amplitudes beyond the leading order in $G$ to claim the precise equivalence between \eqref{RwithX} and \eqref{RwithW}. We will further discuss this issue in the next section.

\section{From black hole spacetime to merger amplitudes}
\label{sec:BHspacetime}
The results in the previous section indicate that the on-shell approach can indeed compute memory effects associated with the gravitational waveform of a merger process. However, despite this optimism, cautious readers may suspect that the approach is valid only to memory types where the radiation frequency is so low that both black holes are effectively treated as point particles. In this section we will show that this is not the case and that the problem of merger in GR can admit an S-matrix interpretation. We will first demonstrate how the final black holes in our Hilbert space --- previously denoted collectively as $X$ --- can be interpreted as modes with proper boundary conditions at the event horizon. The absorption three-point amplitude, responsible for merger, will then be related to these modes. This will naturally introduce a notion of semi-classical S-matrix, which we will use to re-derive the absorption cross section for a massive scalar on a black hole and also the waveform with complete agreement with classical physics.

\subsection{Absorption by a Schwarzschild black hole}
We start by considering a conventional problem in GR, namely the scattering of a massive scalar field in a Schwarzschild background (see e.g.~\cite{Unruh:1976fm, Frolov:1998wf, Glampedakis:2001cx, Bautista:2021wfy})
\begin{align}
\dd s^2=f(r)\dd t^2 -\frac{\dd r^2}{f(r)} - r^2 \dd \Omega^2
\,, \quad
f(r):=1-\frac{2GM}{r}
\,.
\end{align}
A test massive scalar field is governed by the Klein-Gordon equation
\begin{align}
(\nabla_{\mu}\nabla^{\mu}+\mu^2)\varphi = 0
\,.\label{KGeq}
\end{align}
Taking the WKB (or point-particle) limit of the massive scalar field, we can interpret the problem here as a two-body problem with a large mass ratio in GR. Especially, employing the effective-one-body picture~\cite{Buonanno:1998gg, Buonanno:2000ef}, we can think of $M$ and $\mu$ as the total and reduced masses of the two-body system, respectively. This is essentially the same as how a quantum-mechanical potential scattering is obtained from a QFT 2-to-2 scattering. The on-shell approach is of course the QFT-based approach while we can think of the analysis in this section as a QM-based approach. They discuss the same physics in a large mass ratio $\mu \ll M$ but in a different language. We take the latter to put our on-shell approach in contact with a conventional GR problem and a BH spacetime. Note that it is useful not to take the classical limit because the on-shell approach is quantum-based. We keep for a while the wave nature of the particle to see the correspondence between the GR scattering problem and the on-shell approach, and then take the classical limit to reduce the problem into the two-body problem.

\begin{figure}[t]
\centering
\includegraphics[width=\linewidth]{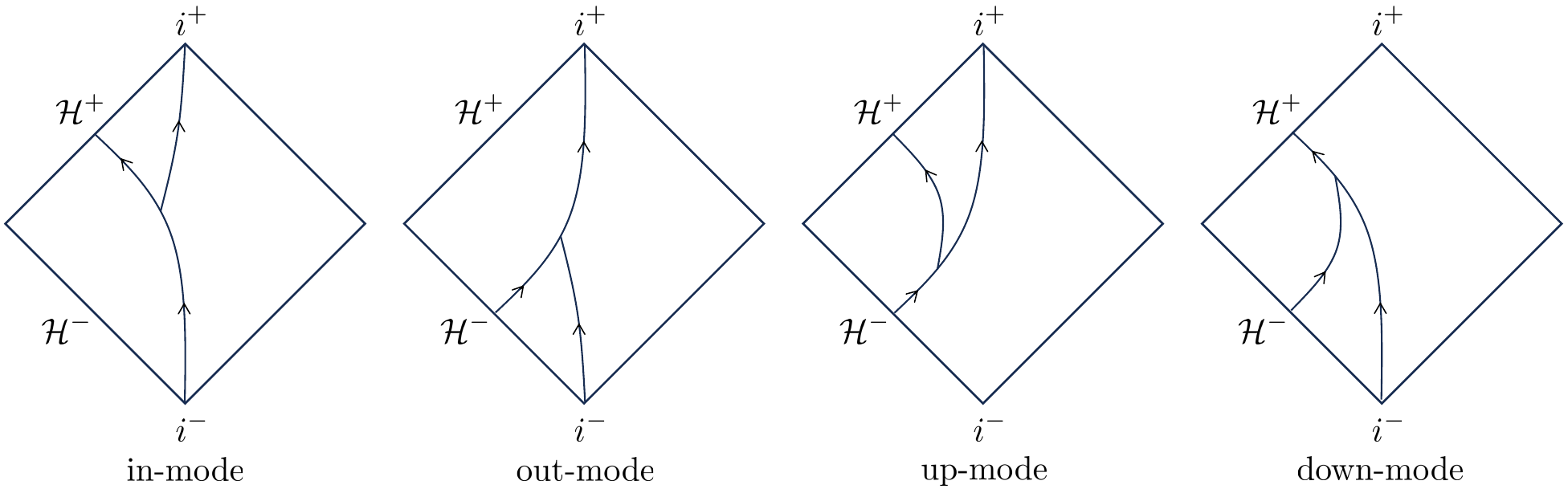}
\caption{Boundary conditions for mode functions where $i^{\pm}$ and $\mathcal{H}^{\pm}$ are past/future timelike infinities and horizons.}
\label{fig:mode}
\end{figure}

Let us introduce the mode functions
\begin{align}
\varphi_{\epsilon \ell m}:=e^{-i\epsilon t} \frac{u_{\ell}(r)}{r} Y_{\ell, m}(\hat{x})
\end{align}
where $\hat{x}$ is the unit spatial vector pointing $(\theta, \phi)$ on the sphere and the radial function $u_{\ell}$ solves
\begin{align}
\left[\frac{\dd^2}{\dd r_*^2}+\epsilon^2-V(r) \right]u_{\ell}(r)=0
\,, \quad V(r):=f(r)\left(\mu^2 + \frac{\ell (\ell+1)}{r^2}  + \frac{2GM}{r^3} \right)
\,,
\label{radial_eq}
\end{align}
with $r_*=r+2GM\log(r/2GM - 1)$ being the tortoise coordinate. The general solution to the radial equation \eqref{radial_eq} is given by a linear combination of any two of the ``in'', ``up'', ``out'', and ``down'' functions. The in and up modes are defined by the purely ingoing boundary condition at the horizon and the purely outgoing boundary condition at the infinity, respectively. The out and down modes are the conjugate of the in and up, $u^{\rm out}_{\ell}=(u^{\rm in}_{\ell})^*, u^{\rm down}_{\ell}=(u^{\rm up}_{\ell})^*$. See Fig.~\ref{fig:mode} for illustration. We use the following normalisation
\begin{align}
u^{\rm in}_{\ell}&\sim 
\begin{cases}
\frac{1}{\sqrt{2\epsilon}} T_{\ell} e^{-i\epsilon r_*}, & r_* \to -\infty\,, \\
\frac{1}{\sqrt{2P}} (R_{\ell}e^{i P r_*} + e^{-i P r_*}), & r_*\to \infty\,,
\end{cases}
\\
u^{\rm up}_{\ell}&\sim 
\begin{cases}
\frac{1}{\sqrt{2\epsilon}} (e^{i\epsilon r_*} - \frac{T_{\ell} R_{\ell}^*}{T_{\ell}^*} e^{-i\epsilon r_*}), & r_* \to -\infty\,, \\
\frac{1}{\sqrt{2P}} T_{\ell} e^{i P r_*},  & r_*\to \infty\,,
\end{cases}
\end{align}
with $P=\sqrt{\epsilon^2-\mu^2}$. $R_{\ell}$ and $T_{\ell}$ are the reflection and transmission coefficients satisfying
\begin{align}
|R_{\ell}|^2 + |T_{\ell}|^2 = 1\,.
\end{align}
$R_{\ell}$ and $T_{\ell}$ can be determined by solving the radial equation, but we are not interested in their concrete forms at this stage. We also add superscripts, in/out/down/up, to specify the boundary conditions of the mode function $\varphi_{\epsilon \ell m}$. The in/up mode functions and the out/down mode functions are related by
\begin{align}
\varphi^{\rm in}_{\epsilon \ell m} &= R_{\ell} \varphi^{\rm out}_{\epsilon \ell m} + T_{\ell } \varphi^{\rm down}_{\epsilon \ell m } 
\,,
\label{in-outdown}
\\
\varphi^{\rm up}_{\epsilon \ell m} &= T_{\ell}  \varphi^{\rm out}_{\epsilon \ell m} - \frac{T_{\ell} R_{\ell}^*}{T_{\ell}^*} \varphi^{\rm down}_{\epsilon \ell m } 
\,.
\label{up-outdown}
\end{align}

We seek a scattering solution consisting of an incident plane wave and an outgoing scattered wave
\begin{align}
\varphi^{+}_{P}\sim e^{-i\epsilon t} \left[ e^{i\vec{P} \cdot \vec{x}} + \frac{f_{P}(\theta)}{r}e^{i P r_*} \right],  \quad r_* \to \infty
\label{asymptotic_in}
\end{align}
under the purely ingoing boundary condition at the horizon where the angle $\theta$ is defined by $\vec{P}\cdot \vec{x} =P r_* \cos\theta$. The in function clearly satisfies the desired boundary conditions. Hence, using the partial wave expansion, we obtain
\begin{align}
\varphi^{+}_{P} &=\sum_{\ell=0}^{\infty} \sum_{m=-\ell}^{\ell}   \frac{4\pi i}{\sqrt{2P}} (-1)^{\ell }  Y^*_{\ell, m}(\hat{P})  \varphi^{\rm in}_{\omega \ell m}(t,r,\hat{x})
\end{align}
with
\begin{align}
f_P(\theta) &= \sum_{\ell=0}^{\infty} (2\ell +1) \frac{\eta_{\ell}e^{2i\delta_{\ell}}-1}{2iP} P_{\ell}(\hat{P}\cdot \hat{x})
\,, \quad 
\eta_{\ell} e^{2i\delta_{\ell}} = (-1)^{\ell +1} R_{\ell}
\,,
\end{align}
where $P_{\ell}$ are the Legendre polynomials and $\hat{P}=\vec{P}/|P|$. 
The function $f_P(\theta)$ is known as the scattering amplitude in the time-independent scattering theory. For our purpose, it is more convenient to take the time-dependent scattering theory (see e.g.~\cite{weinberg2015lectures}) by introducing in/out states. As the asymptotic form \eqref{asymptotic_in} corresponds to the Lippmann-Schwinger equation with $+i\varepsilon$, we can interpret $\varphi^+_{P}$ as the in state with momentum $\vec{P}$. The out state is defined by the solution whose asymptotic form consists of a plane wave and an ingoing spherical wave
\begin{align}
\varphi^{-}_{P}&\sim e^{-i\epsilon t} \left[ e^{i\vec{P} \cdot \vec{x}} + \frac{f^*_{-P}(\theta)}{r}e^{-i P r_*} \right],   \quad r_* \to \infty
\,.
\end{align}
By using the partial wave expansion together with $f_{-P}(\theta)=f_{P}(\pi-\theta)$ and $P_{\ell}(-z)=(-1)^{\ell}P_{\ell}(z)$, one can recognise that this asymptotic form corresponds to the out mode 
\begin{align}
\varphi^{-}_{P} = -\sum_{\ell=0}^{\infty}\sum_{m=-\ell}^{\ell}   \frac{4\pi i}{\sqrt{2P}}  Y^*_{\ell, m}(\hat{P}) \varphi^{\rm out}_{\epsilon \ell m}
\,.
\end{align}
In the Schwarzschild spacetime, we have two more mode functions, up and down, which are the waves emitted from the horizon in the past and absorbed into the horizon in future, respectively. Is there any on-shell interpretation for these states? First of all, let's notice that an observer on $i^{-}$ does not observe the up mode because this mode doesn't have an ingoing wave at $r_* \to \infty$. The down mode is similar: an observer on $i^{+}$ cannot see the outgoing wave. From the viewpoint of such observers, hence, the up and down modes can be thought of not as propagating waves but as states belonging to the background. These are precisely the $X$-states we have been discussing in the previous sections. 
Therefore, we regard the up/down as the in/out states of $X$ and write
\begin{align}
X^{+}_{\epsilon \ell m} = \varphi^{\rm up}_{\epsilon \ell m}
\,, \quad
X^{-}_{\epsilon \ell m} = \varphi^{\rm down}_{\epsilon \ell m}
\,.
\end{align}

Motivated by the time-dependent scattering theory of quantum mechanics and \cite{Verlinde:1991iu, Kabat:1992tb, Fabbrichesi:1993kz}, we define the S-matrix of this problem by inner products of the mode functions. The natural candidate for the inner product is
\begin{align}
(\varphi_1, \varphi_2):=i \int_{\Sigma} \dd \Sigma^{\mu}(\varphi^*_1 \partial_{\mu} \varphi_2 - \varphi_2 \partial_{\mu} \varphi^*_1 )
\,,
\end{align}
where $\dd \Sigma^{\mu}$ is the future-directed vector of the Cauchy surface $\Sigma$. The inner product does not depend on the particular choice of the Cauchy surface. For instance, we can choose the future Cauchy surface where the out and down modes have values on only one of $\mathcal{H}^+$ or $i^+$, so they are orthogonal. The inner product of the same state is 
\begin{align}
(\varphi^A_{J'}, \varphi^A_J) = \hat{\delta}_{J',J}
\end{align}
and other non-vanishing inner products are
\begin{align}
(\varphi^{\rm out}_{J'}, \varphi^{\rm in}_{J}) &=R_{\ell} \hat{\delta}_{J',J} \,, \quad (\varphi^{\rm down}_{J'}, \varphi^{\rm in}_{J}) = T_{\ell} \hat{\delta}_{J',J} \,, \\
(\varphi^{\rm out}_{J'}, \varphi^{\rm up}_{J}) &=T_{\ell} \hat{\delta}_{J',J} \,, \quad (\varphi^{\rm down}_{J'}, \varphi^{\rm up}_{J}) = -\frac{T_{\ell} R^*_{\ell}}{T^*_{\ell} } \hat{\delta}_{J',J} 
\,,
\end{align}
where $J=\{\epsilon \ell m\}$ is the collective index and 
\begin{align}
\hat{\delta}_{J', J}:= \hat{\delta}(\epsilon' - \epsilon)\delta_{\ell', \ell} \delta_{m',m}
\,.
\end{align}
Then, the S-matrix is given by
\begin{align}
\braket{\alpha| S |\beta}:=
\begin{pmatrix}
(\varphi^-_{P'}, \varphi^+_P) & (\varphi^-_{P'}, X^+_J)  \\
(X^-_{J'}, \varphi^+_P) & (X^-_{J'}, X^+_J )
\end{pmatrix}
\,,
\end{align}
where each components are computed as
\begin{align}
\braket{P'| S |P} &=\hat{\delta}_{\Phi}(P'-P)  + 4\pi i \hat{\delta}(\epsilon'-\epsilon) f(P', P)
\,, 
\label{SPP}\\
\braket{J'| S |P}&= 4\pi i \hat{\delta}(\epsilon'-\epsilon) \frac{T_{\ell'}}{\sqrt{2P }} (-1)^{\ell'}Y^*_{\ell', m'}(\hat{P}) 
\,, 
\label{SJP}\\
\braket{P'| S |J} &=4\pi i \hat{\delta}(\epsilon'-\epsilon) \frac{T_{\ell}}{\sqrt{2P}} Y_{\ell , m}(\hat{P}')
\,, 
\label{SPJ}\\
\braket{J'| S |J} &=-\frac{T_{\ell} R_{\ell}^*}{T_{\ell}^*} \hat{\delta}_{J',J}
\,,
\end{align}
with $\hat{\delta}_{\Phi}(P'-P)=2\epsilon \hat{\delta}^{(3)}(\vec{P}'-\vec{P}) $ and
\begin{align}
f(P', P)&:= \sum_{\ell=0}^{\infty} (2\ell +1) \frac{\eta_{\ell}e^{2i\delta_{\ell}}-1}{2i P} P_{\ell}(\hat{P}'\cdot \hat{P})
\,.
\end{align}
The S-matrix we have introduced is unitary on the full Hilbert space
\begin{align}
\sum_{\gamma} \braket{\alpha| S^{\dagger} |\gamma} \braket{\gamma| S |\beta} =
\int_P \braket{\alpha| S^{\dagger} |P} \braket{P| S |\beta} + \int_J  \braket{\alpha| S^{\dagger} |J} \braket{J| S |\beta} = \braket{\alpha|\beta} 
\end{align}
provided that the phase-space measure and the normalisation of the states are given by
\begin{align}
\int_P :=\int \dd \Phi(P) = \int \frac{\hat{\dd}^3 P}{2\epsilon}
\,, \quad 
\int_J := \sum_{\ell=0}^{\infty}\sum_{m=-\ell}^{\ell} \int_0^{\infty} \hat{\dd}\epsilon
\end{align}
and
\begin{align}
\braket{\alpha|\beta} 
:=
\begin{cases}
\hat{\delta}_{\Phi}(P'-P) &\alpha=P',~\beta=P\,, \\
\hat{\delta}_{J',J} &\alpha= J',~\beta=J\,, \\
0 & \text{otherwise}\,.
\end{cases}
\end{align}

Since the state $P$ is understood as the relative motion of the two-body system and the state $J$ is the fusion state with the background, they should be matched with the two-particle state and the one-particle state in QFT, respectively. Indeed, one can see that the kinematic dependence of $\bra{J'}S\ket{P}$ shown in \eqref{SJP} is the same as the massive three-point amplitude presented in \eqref{3pt_up}.\footnote{Needless to say, $f(P,P')$ can be matched with the $12\to 12$ scattering amplitude in \eqref{12->12}.} However, the direct comparison between \eqref{3pt_up} and \eqref{SJP} may be subtle because the centre-of-mass motion is not included in the latter and there are ambiguities of the normalisation and the phase. Hence, the easiest way is to look at the observable namely the absorption cross-section
\begin{align}
\dd \sigma^{\rm abs} &:= \frac{V}{vT} \frac{|\braket{J|S|P}|^2}{\braket{J|J}\braket{P|P}} \dd \Pi_J
\,.
\end{align}
Here, $V$ is the spatial volume, $T$ is the time of the reaction, $v=P/\epsilon$ is the initial velocity, and $\dd \Pi_J$ is the phase-space measure of the final state $J$ with the normalisation $\int \dd \Pi_J = 1$. Noting $\hat{\delta}(\epsilon-\epsilon)=T$ and $\hat{\delta}^{(3)}(\vec{P}-\vec{P})=V$, we obtain
\begin{align}
\sigma^{\rm abs} &=\sum_{\ell=0}^{\infty}\sum_{m=-\ell}^{\ell} \int_0^{\infty} \dd \epsilon' \delta(\epsilon-\epsilon') \frac{4\pi^2}{P^2} |T_{\ell}|^2Y^*_{\ell, m}(\hat{P}) Y_{\ell ,m}(\hat{P})
\nn
&=\sum_{\ell=0}^{\infty}\sum_{m=-\ell}^{\ell} \int_{(m_1+m_2)^2}^{\infty} \dd m_X^2 \delta(s-m_X^2) \frac{16\pi^2 m_X^2}{\lambda(m_1^2,m_2^2,m_X^2)} |T_{\ell}|^2 Y^*_{\ell, m}(\hat{P}) Y_{\ell ,m}(\hat{P})
\end{align}
where $m_X^2=M^2+2M\epsilon'$ and $s=M^2+2M\epsilon$, and it should be recalled that $M=m_1+m_2$ is the total mass and $\lambda(m_1^2,m_2^2,m_X^2)=4E^2P^2$. We compare this with \eqref{abs_cross-section} by substituting \eqref{3pt_up}, which is 
\begin{align}
\sigma^{\rm abs} &=\sum_{\ell=0}^{\infty}\sum_{m=-\ell}^{\ell} \int_{(m_1+m_2)^2}^{\infty} \dd m_X^2 \delta(s-m_X^2) 
\rho_{\ell}(m_X^2) |g_{\ell}|^2 
\nn
&\qquad 
\times \frac{4\pi^2}{2\ell +1} \frac{(\ell!)^2}{(2\ell)!} \frac{[m_X \lambda^{1/2}(m_1^2,m_2^2,m_X^2)]^{\ell}}{\lambda^{1/2}(m_1^2,m_2^2,m_X^2)} Y^*_{\ell, m}(\hat{P}) Y_{\ell ,m}(\hat{P})
\,.
\label{abs_cross_GR}
\end{align}
\newline
Therefore, we obtain the matching condition for the large mass ratio as follows:
\begin{align}
\rho_{\ell} |g_{\ell}|^2 = \frac{4(2\ell+1)|T_{\ell}|^2 }{\lambda^{1/2}(m_1^2,m_2^2,m_X^2)}   \frac{(2\ell)!}{(\ell!)^2} \frac{m_X^2 }{[m_X \lambda^{1/2} (m_1^2,m_2^2,m_X^2)]^{2\ell}} 
\,.
\label{matching_q}
\end{align}
\newline
Note that the discussions so far are solely based on the boundary conditions of the Klein-Gordon equation \eqref{KGeq} and the specific form of the Schwarzschild spacetime has not been used, which highlight a very general aspect of our approach. The information of the background appears only in the transmission coefficient $T_{\ell}$. Hence, we can apply a similar analysis for other black hole metrics such as the effective-one-body metric~\cite{Buonanno:1998gg, Buonanno:2000ef} as long as the causal structure in the exterior region is the same. We also note that we have not taken the classical limit of the relative motion while the centre-of-mass motion (the background spacetime) is classical. So, the matching condition \eqref{matching_q} is a quantum-level matching.

We now take the classical limit to describe the classical absorption cross-section. In this limit, eq.~\eqref{abs_cross_GR} is simplified to be
\begin{align}
\sigma^{\rm abs} = \sum_{\ell =0}^{\infty} \frac{\pi(2\ell+1)}{P^2} |T_{\ell}|^2 
\,.
\end{align}
By restoring $\hbar$, the radial equation in the $\hbar \to 0$ limit simplifies to
\begin{align}
\left[\hbar^2 \frac{\dd^2}{\dd r_*^2}+\epsilon^2-V_{\rm cl} \right]u_{\ell}(r)=0\,, \qquad V_{\rm cl}=f(r)\left(\mu^2 + \frac{\ell^2}{r^2}\right)
\,,
\label{radial_cl}
\end{align}
which can be solved by the WKB approximation,
\begin{align}
u_{\ell}(r)=\frac{c_1}{\sqrt{2P_r(r)}}e^{+\frac{i}{\hbar} \int dr'_* P_r(r'_*)}   
+\frac{c_2}{\sqrt{2P_r(r)}}e^{- \frac{i}{\hbar} \int dr'_* P_r(r'_*)}
\,, \quad
P_r(r):=\sqrt{\epsilon^2-V_{\rm cl}(r)}
\,.\label{WKB}
\end{align}
Using this, it is straightforward to check that the transmission coefficient $T_{\ell}$ is unity for $\ell$ such that $\epsilon-V_{\rm cl}(r)$ is never zero outside the horizon, and is zero otherwise. The angular momentum defining this transition is denoted as critical angular momentum and it is given by
\begin{align}
L_c^2 = (GM\epsilon)^2  \left\{ \frac{[8(1-v^2)]^3}{4[1-4v^2+(1+8v^2)^{1/2}] [3-(1+8v^2)^{1/2}]^2} \right\}
\end{align}
where the quantity in the curly brackets takes values between 16 and 27 for $0\leq v\leq 1$. The typical size is $L_c \sim GM\epsilon \sim GM\mu=Gm_1m_2$. Then, the absorption cross-section in the classical limit is~\cite{Unruh:1976fm}
\begin{align}
 \sigma^{\rm abs} = \frac{\pi L_c^2}{P^2}=\frac{\pi (GM)^2}{v^2}\left\{ \frac{[8(1-v^2)]^3}{4[1-4v^2+(1+8v^2)^{1/2}] [3-(1+8v^2)^{1/2}]^2} \right\} \quad {\rm as} \quad \hbar\to 0
 \,.
\end{align}
One can also obtain the same absorption cross-section by using the geodesic motion. In classical physics, the absorption cross-section is the area where a particle crossing the region falls into the black hole
\begin{align}
\sigma_{\rm cl}^{\rm abs} := \int_0^{b_c} \dd b\, 2\pi b =\pi b_c^2
\,.
\label{cross_cl}
\end{align}
Since the radial motion of the geodesic is given by the same effective potential as \eqref{radial_cl}, the critical impact parameter is $b_c=L_c/P$, giving the same result. In the relativistic limit $v\to 1$, the absorption cross-section approaches the area of the BH shadow $27\pi (GM)^2$. Instead of using \eqref{matching_q}, we can use the classical prediction \eqref{cross_cl} and the fact that a BH has to be a complete absorber as a classical-level matching condition. We can then determine the on-shell three-point without referring to quantum physics.

\subsection{Radiation from infalling particle}

We have seen that the absorption of a massive particle by a black hole naturally admits an S-matrix interpretation. Let's now turn to the problem of radiation emitted during the process. 
To avoid any unimportant complications we will focus on a simplified model given by a toy model of a massless scalar field $h$ by introducing the following coupling to the massive field $\varphi$
\begin{align}
\mathcal{L}_{\rm int}=\frac{\lambda}{2} h \varphi^2
\,.
\label{interaction}
\end{align}
This is a simplified version of the gravitational coupling $\frac{\kappa}{2} h_{\mu\nu}T^{\mu\nu}$, but it captures all the essential points. We are thus interested in the problem of a test particle, described by a localised wavepacket of the massive scalar field $\varphi$, falling into a Schwarzschild BH while emitting a massless scalar wave $h$ via \eqref{interaction}. 

We first solve this problem by using the KMOC formalism but in a quantum-mechanical setup. The initial state is given by
\begin{align}
\ket{\Psi}=\int_P \phi(P) e^{-i \vec{b} \cdot \vec{P}}\ket{P} \, ,
\label{initial_relative}
\end{align}
where $\phi(P)$ is the wavefunction of the relative motion. The waveform of $h$ at the future null infinity is then
\begin{align}
h&= \frac{1}{4\pi  r} \int^{\infty}_0 \hat{\dd} \omega e^{-i\omega u} W +{\rm c.c.} 
\label{waveform_h}
\end{align}
with the spectral waveform
\begin{align}
iW:=\bra{\Psi}S^{\dagger} a_k S\ket{\Psi} =\int_{P,P'}\phi^*(P')\phi(P) e^{i \vec{b} \cdot (\vec{P}'-\vec{P})} \bra{P'}S^{\dagger}a_k S\ket{P}
\,.
\label{W_h}
\end{align}
Using the completeness relation and keeping only the leading terms in $\lambda$, we find
\begin{align}
\bra{P'}S^{\dagger}a_k S\ket{P} = \int_{P''}\bra{P'}S^{\dagger}\ket{P''}\bra{P'';k}S\ket{P}  + \int_{J}\bra{P'}S^{\dagger}\ket{J}\bra{J;k}S\ket{P} + \mathcal{O}(\lambda^2)
\,. 
\label{PSaSP}
\end{align}
Since $\bra{P'}S^{\dagger}\ket{P''}$ and $\bra{P'}S^{\dagger}\ket{J}$ are given in \eqref{SPP} and \eqref{SJP} at the zeroth order in $\lambda$, the remaining task is the computation of the amplitudes for $P\to J,k $ and $P\to P'',k$ at linear order in $\lambda$. It has been argued that tree-level amplitudes on curved spacetimes, at least in the semi-classical sense, may be obtained by evaluating the on-shell action known as the perturbiner approach~\cite{Arefeva:1974jv,Jevicki:1987ax,Selivanov:1997aq, Mizera:2018jbh, Cho:2023kux,Aoki:2024bpj}. We adopt this approach with the following corrections due to the presence of the horizon. First, the in/out states of $P, J, k$ are given by the in/out mode functions introduced above where the in/out modes of $h$ are denoted by $h^{\pm}_{k}$ which satisfy the same ingoing/outgoing boundary conditions as $\varphi^{\pm}_P$. Second, we restrict the domain of integral to the exterior region of the Schwarzschild spacetime $r>r_S=2GM$. The first one may be naturally understood by the discussions in the previous subsection while the second one is assumed because the distant observer can only access the exterior region. Then, by following the perturbiner, the S-matrix is computed as
\begin{align}
\bra{P'';k}S\ket{P} &= \lambda \int_{r>r_S} \dd^4 x \sqrt{-g} (h^-_k)^* (\varphi^-_{P''})^* (\varphi^+_P)\,, \label{P->Pk} \\
\bra{J;k}S\ket{P} &=\lambda \int_{r>r_S} \dd^4 x \sqrt{-g} (h^-_k)^* (X^-_J)^* (\varphi^+_P)\,, \label{P->Jk}
\end{align}
and the waveform \eqref{waveform_h} is obtained accordingly. We shall not either perform the integrals or take the classical (point-particle) limit because \eqref{PSaSP} with \eqref{P->Pk} and \eqref{P->Jk} are sufficient to compare the KMOC formalism with the classical calculations shown below.

Next, we compute the waveform by solving the classical equations of motion:
\begin{align}
(\Box +\mu^2)\varphi=\lambda h \varphi\,, \quad \Box h=\frac{\lambda}{2} \varphi^2
\,.
\end{align}
When the coupling constant $\lambda$ is small, we can ignore the radiation reaction to $\varphi$, allowing us to approximate our equations of motion by
\begin{align}
(\Box+\mu^2) \varphi = 0 \,, \quad \Box h = \frac{\lambda}{2} \varphi^2
\,.
\end{align}
Hence, the solution of $\varphi$ is given by a linear combination of the mode functions introduced in the previous subsection at the leading order. We are interested in the solution describing an incident wavepacket at the past infinity with the impact parameter $\vec{b}$ under the absorbing boundary condition on the horizon. Hence, the classical solution of $\varphi$ is
\begin{align}
\varphi_{\rm cl} = \int \dd\Phi(P) \phi(P) e^{-i\vec{b}\cdot \vec{P}} \varphi^+_{P} + {\rm c.c.}
\end{align}
where $\phi(P)$ is the wavefunction of this classical wavepacket mimicking a point particle, which we may identify with the wavefunction in \eqref{initial_relative} in the classical limit. To find the solution $h$, we perform the Fourier transformation and expand it into the spherical basis
\begin{align}
h_{\rm cl}= \sum_{\ell, m} \int_{0}^{\infty} \hat{\dd} \omega e^{-i\omega t} \frac{v_{\omega \ell m}(r)}{r} Y_{\ell, m}(\hat{x})   + {\rm c.c.}
\end{align}
where the equation of motion for the radial function is
\begin{align}
\left[ \frac{\dd^2}{\dd r_*^2} + \omega^2 - U(r) \right] v_{\omega \ell m}(r)  = S_{\omega \ell m}(r)
\label{radial_h}
\end{align}
with the source
\begin{align}
S_{\omega \ell m} &=-\lambda r f(r) \int \dd t \dd^2\hat{x} e^{i\omega t} Y^*_{\ell, m}(\hat{\bm x}) \varphi_{\rm cl}^2 
\,.
\end{align}
The potential $U(r)$ is given by \eqref{radial_eq} while setting $\mu=0$. The source term is composed of two different frequencies, the high-frequency modes $\omega \sim \pm 2\mu$ coming from $\varphi_{P'}^+ \varphi_P^+, (\varphi_{P'}^+)^* (\varphi_{P}^+)^*$ and the low-frequency modes $\omega \ll \mu$ from $(\varphi_{P'}^+)^*(\varphi_{P}^+)$. For a massive particle with an astrophysical mass $\mu \gg \Mpl$, the emission of the former should be highly suppressed and is not of interest to us. Hence, we only consider the low frequencies in which the source takes the form
\begin{align}
S_{\omega \ell m}&= \int_{P,P'}\phi^*(P') \phi(P) e^{i\vec{b}\cdot(\vec{P}'-\vec{P})} \int \dd t \dd^2\hat{x}\,  \lambda r f(r) e^{i\omega t} Y^*_{\ell, m}(\hat{x})  (\varphi^+_{P'})^* \varphi^+_{P} \,.
\end{align}

We solve the radial equation \eqref{radial_h} under the boundary condition that no waves are injected from infinity and the horizon. The appropriate Green's function for such boundary conditions is
\begin{align}
G_{\ell}(r_*, r_*') =\frac{1}{2i \omega A_{\ell}^{\rm in}} \left[ \theta(r_*-r_*') \hat{v}^{\rm in}_{\ell} (r'_*) \hat{v}_{\ell}^{\rm up}(r_*) + \theta(r'_*-r_*) \hat{v}_{\ell }^{\rm in}(r_*) \hat{v}_{\ell}^{\rm up}(r'_*) \right]
\,,
\end{align}
where the hatted functions are defined to satisfy the following boundary conditions and the normalisations to follow the standard convention of the black hole perturbation (e.g.~\cite{maggiore2018gravitational}):
\begin{align}
\hat{v}^{\rm in}_{\ell}&\sim
\begin{cases}
e^{-i\omega r_*}, & r_* \to -\infty\,, \\
A^{\rm out}_{\ell}e^{i\omega r_*} + A^{\rm in}_{\ell} e^{-i\omega r_*}, & r_*\to \infty\,,
\end{cases}
\\
\hat{v}^{\rm up}_{\ell}&\sim
\begin{cases}
B^{\rm out}_{\ell}e^{i\omega r_*} + B^{\rm in}_{\ell} e^{-i\omega r_*}, & r_* \to -\infty\,, \\
e^{i\omega r_*},  & r_*\to \infty\,.
\end{cases}
\end{align}
We are particularly interested in the asymptotic form of $h_{\rm cl}$ at the future null infinity. Therefore, the solution to (\ref{eq:radial-rad}) at $r_* \to \infty$ is
\begin{align}\label{eq:radial-rad}
v_{\omega \ell m} 
&= e^{i \omega r_*} \frac{1}{2i \omega A_{\ell}^{\rm in}} \int^{\infty}_{-\infty} \dd r_*' \hat{v}^{\rm in}_{\ell}(r_*') S_J(r_*') 
\nn
&=e^{i \omega r_*} \int_{P,P'} \phi^*(P') \phi(P) e^{i\vec{b}\cdot(\vec{P}'-\vec{P})} \int_{r>r_S} \dd^4x \sqrt{-g} \frac{\lambda}{i \sqrt{2\omega} }  (h^{\rm out}_{\omega \ell m})^* (\varphi^+_{\bm{k}'})^* \varphi^+_{\bm k} .
\end{align}
Here, we have used that the in mode is the complex conjugate of the out mode $\hat{v}^{\rm in}_{\ell}=(\hat{v}^{\rm in}_{\ell })^*$ and the black-hole-perturbation normalisation and the quantum-mechanical normalisation are related by $\hat{v}^{\rm in}_{\ell }=\sqrt{2\omega} A_{\ell}^{\rm in} v^{\rm in}_{\ell} $. In addition, one should recall that the spherical mode function of $h$ is
\begin{align}
    h^{\rm out}_{\omega \ell m}=e^{-i\omega t}\frac{v^{\rm out}_{\ell}}{r}Y_{\ell,m}
    \,.
\end{align}
Then, using the out state of $h$ with the momentum $k$,
\begin{align}
    h^-_k = -\sum_{\ell,m} \frac{4\pi i}{\sqrt{2\omega}}Y^*_{\ell,m}(\hat{k})h^{\rm out}_{\omega \ell m}
    \,,
\end{align}
the classical waveform is computed as
\begin{align}
    h_{\rm cl}=\frac{1}{4\pi r}\int_0^{\infty}\hat{\dd}\omega e^{-i\omega u}W_{\rm cl}+{\rm c.c.},
\end{align}
with
\begin{align}
    iW_{\rm cl}=\int_{P,P'}\phi^*(P')\phi(P)e^{i\vec{b}\cdot(\vec{P}'-\vec{P})}\left[\lambda \int_{r>r_S}\dd^4 x \sqrt{-g} (h^-_k)^* (\varphi^+_{P'})^* \varphi^+_P \right]
    \,.
    \label{classical_W}
\end{align}

The classical waveform already takes the same form as the one computed by KMOC \eqref{waveform_h} and \eqref{W_h} under the identification
\begin{align}
    \bra{P'}S^{\dagger}a_k S\ket{P}= \lambda \int_{r>r_S}\dd^4 x \sqrt{-g} (h^-_k)^* (\varphi^+_{P'})^* \varphi^+_P 
    \,. \label{inclusive}
\end{align}
To further clarify the relationship to scattering amplitudes, we should recall that the in state is related to the out and down states by \eqref{in-outdown}, giving
\begin{align}
\varphi_{P}^+= \sum_{\ell,m} \frac{4\pi i}{\sqrt{2P}} (-1)^{\ell} Y_{\ell,m}^*(\hat{\bm k}) (R_{\ell} \varphi^{\rm out}_J + T_{\ell} \varphi^{\rm down}_J)
\,.
\end{align}
The reflection and transmission coefficients are related to the elastic and inelastic channels of the S-matrix as shown in \eqref{SPP} and \eqref{SJP}, leading to
\begin{align}
    \sum_{\ell,m} \frac{4\pi i}{\sqrt{2P}} (-1)^{\ell} Y_{\ell,m}^*(\hat{P}) R_{\ell} \varphi^{\rm out}_J &=\int_{P'} \bra{P'}S\ket{P} \varphi^{-}_{P'}
    \,, \\
    \sum_{\ell,m} \frac{4\pi i}{\sqrt{2P}} (-1)^{\ell} Y_{\ell,m}^*(\hat{P}) T_{\ell} \varphi^{\rm down}_J
&=\int_J \bra{J}S\ket{P} X^-_J
\,.
\end{align}
As a result, using \eqref{P->Pk} and \eqref{P->Jk}, we obtain
\begin{align}
    \lambda \int_{r>r_S}\dd^4 x \sqrt{-g} (h^-_k)^* (\varphi^+_{P'})^* \varphi^+_P 
    &=\int_{P''} \bra{P'}S^{\dagger}\ket{P''} \left[ \lambda \int_{r>r_S}\dd^4 x \sqrt{-g} (h^-_k)^* (\varphi^-_{P''})^* \varphi^+_P \right]
    \nn
    &+\int_{J} \bra{P'}S^{\dagger}\ket{J} \left[ \lambda \int_{r>r_S}\dd^4 x \sqrt{-g} (h^-_k)^* (X^-_{J})^* \varphi^+_P \right]
    \nn
    &=\int_{P''} \bra{P'}S^{\dagger}\ket{P''}\bra{P'';k}S\ket{P}
    +\int_J \bra{P'}S^{\dagger}\ket{J}\bra{J;k}S\ket{P}
\end{align}
which is exactly \eqref{PSaSP}. This complete showing that the classical problem of a wave emission from an infalling particle indeed admits an S-matrix interpretation, given that the S-matrix on the Schwarzschild spacetime is computed by the on-shell action. The classical solution $h_{\rm cl}$ contains all important features of the BH merger from the inspiral to the ringdown at the order we are working on, so the amplitude \eqref{P->Jk} should also have all these information.\footnote{We again emphasise that our discussion relies only on the boundary conditions of the mode functions so the concrete form of the spacetime metric is not so important. }

We close this section with two remarks. First, we notice that the object $\bra{P'}S^{\dagger}a_k S\ket{P}$, called the inclusive amplitudes in \cite{Caron-Huot:2023vxl}, admits a simple representation \eqref{inclusive}. In fact, the insertion of the mode functions in \eqref{inclusive} is naturally understood by the fact that the inclusive amplitude is represented by~\cite{Caron-Huot:2023vxl}
\begin{align}
    \bra{P'}S^{\dagger}a_k S\ket{P} = \bra{0}a^{\rm in}_{P'}a^{\rm out}_k (a^{\rm in}_P)^{\dagger}\ket{0}
    \,,
\end{align}
differently from the scattering amplitude $\bra{P';k}S\ket{P}=\bra{0}a^{\rm out}_{P'}a^{\rm out}_k (a^{\rm in}_P)^{\dagger}\ket{0}$. As studied in \cite{Caron-Huot:2023vxl, Caron-Huot:2023ikn}, it would be useful to explore further relations between the scattering and inclusive amplitudes and to investigate an efficient way to compute the latter. Second, we note that the two representations of the momentum radiated \eqref{RwithX} and \eqref{RwithW} can be shown to be the same at the linear order in $\lambda$ in the classical limit. It can be seen by noting that from \eqref{R_transform}, the two are the same at the linear order in $\lambda$ if
\begin{align}
    \bra{P}S^{\dagger}a_k S\ket{J} =  \lambda \int_{r>r_S}\dd^4 x \sqrt{-g} (h^-_k)^* (\varphi^+_{P})^* X^+_J =0
    \label{inclusive_J}
\end{align}
in the classical limit. We apply the WKB approximation for the solutions of the massive mode functions \eqref{WKB}. Then, for the scattering kinematics $(\ell > L_c)$, the integral is zero because $\varphi^+_P$ and $X^+_J$ take non-zero values only outside or inside of the angular momentum barrier of the potential. For the merger kinematics $(\ell<L_c)$, the radial mode functions satisfy $u^{\rm in}_{\ell}=u^{\rm down}_{\ell}=(u^{\rm up}_{\ell})^*$. This implies that the radial integral of \eqref{inclusive_J} rapidly oscillates and should vanish in the classical limit. All in all, we end up with \eqref{inclusive_J} for both kinematics, showing the equivalence between \eqref{RwithX} and \eqref{RwithW} at $\mathcal{O}(\lambda)$.

\section{Conclusion and outlook}
In this paper, we have initiated a program aimed at describing classical black hole mergers through on-shell scattering amplitudes. The key idea is to treat black holes as particles, with their mergers viewed as a fusion process in particle physics, characterized by massive three-point amplitudes. Modern on-shell methods unambiguously determine the amplitude for the merger of two Schwarzschild black holes, except for overall coupling constants. These are then fixed to reflect a fundamental property of black holes: nothing can escape the event horizon. In amplitude terms, black hole scattering must involve complete absorption when two black holes come close enough to form a new one. Therefore, the coupling between three black holes reaches the maximum value allowed by unitarity. We have shown that the massive three-point amplitudes determined in this way correctly give the classical momentum and angular momentum conservations by taking the classical limit equipped with the KMOC formalism \cite{Kosower:2018adc} and the use of coherent spin states \cite{Aoude:2021oqj}.

The massive three-point amplitude we have introduced in (\ref{M3pt}) serves as the building block for computing higher-point amplitudes responsible for gravitational wave emissions. As another application, we have computed the tree-level four-point amplitude, describing two incoming Schwarzschild black holes and one outgoing Kerr black hole and graviton, both minimally coupled.
 We observed that our four-point amplitude is factorized in the manner of the soft theorem, with the soft factor yielding the classical gravitational waveform of the spin memory type, valid at all orders in the classical spin. This re-establishes the well-known relation between the soft theorem and the gravitational memory but in a purely on-shell way for merger processes. 

We have further examined how dynamics on a black hole spacetime can be reinterpreted in the S-matrix language by comparing the amplitude approach with the conventional perturbative approach around a black hole background. From this perspective, the comparison can be understood as whether we solve the same problem using a quantum-field-theoretic approach or a quantum-mechanical one. The latter is limited to the large mass ratio but is useful because the equations of motion are simplified and can be solved semi-analytically. In particular, the amplitude \eqref{P->Pk} may be thought of as the all-order $G$ resummation of a five-point amplitude but at leading order in the mass ratio as in \cite{Adamo:2023cfp}. However, a relevant contribution is also \eqref{P->Jk} --- unnoticed in \cite{Adamo:2023cfp} --- which can be thought of as a resummed four-point amplitude describing the merger process, namely our $\bra{X,\ell; k}S\ket{p_1;p_2}$. 

Needless to say, the results in this paper only represent a very first step toward the on-shell description of black hole mergers. The result in Sec.~\ref{sec:radiation} does not require any assumptions on the mass ratio although it is linear in $G$. On the other hand, Sec.~\ref{sec:BHspacetime} only applies for the large mass ratio but is to all orders in $G$. We should combine these complementary approaches to achieve this goal. An obvious direction is the inclusion of loop corrections and the (would-be eikonal) resummations of on-shell amplitudes. Our tree-level amplitude has no graviton exchange between massive particles, implying no gravitational force between the incoming black holes. In other words, before and after the merger our black holes are free and gravitational radiation is emitted only due to the change of the initial and final momenta according to the conservation law, not by the gravitational acceleration of the black holes. This is precisely the gravitational memory~\cite{Braginsky:1987kwo} which explains why our waveform is a pure memory type. On the other hand, the information of bulk dynamics is surely in $\bra{J;k}S\ket{P}$. Hence, we can naturally expect that the dynamics appear at a loop level where gravitons are exchanged between black holes.

Yet, just adding gravitons should be insufficient to describe black hole physic because it lacks the feature of the event horizon. The absorption effect by the horizon can be taken into account by mass-changing (unequal-mass) three-point amplitudes~\cite{Jones:2023ugm,Aoude:2023fdm,Chen:2023qzo}; that is, a black hole will change to another state with a different mass by absorbing a graviton. The absorption effect is particularly important in the high-frequency region $\omega \gtrsim 1/Gm$, namely the ringdown. This makes sense also from the EFT point of view because the high-frequency modes can resolve UV resonance states, the quasi-normal modes in the context of black holes. Said another way, new states should be required to describe the ringdown and the mass-changing amplitudes may be regarded as the interactions with such new states. Combining a resummation of PM amplitudes with mass-changing amplitudes could serve as an on-shell analogue of the combination of the EOB resummation of PN expansion and the black hole perturbations.

There are also other directions to develop. For instance, one can consider collisions of Kerr black holes rather than Schwarzschild. The three-point amplitudes of massive spin-0 particles and a massive spin-$s$ are uniquely fixed while three-point amplitudes for all spinning legs are not unique~\cite{Arkani-Hamed:2017jhn}. We should thus ask what three-point amplitude corresponds to this type of collision. Similarly, higher-point amplitudes such as a process of $N$ spin-0 particles to a spin-$s$ are not unique, which might be regarded as a gravitational collapse of matters. It would be also intriguing to relate our results for the sub-subleading soft-graviton theorem with recent discussions on large gauge transformations from an on-shell perspective \cite{Elkhidir:2024izo}.
Another interesting direction would be studying merger problems in the Yang-Mills theory and asking if there exists a double-copy structure, something like a $\sqrt{\text{BH-merger}}$ in Yang-Mills. Furthermore, one might get a new insight into the quantum physics of black holes such as Hawking radiation and information loss paradox by applying the on-shell approach to black holes. They would all help to expose black holes to the modern on-shell picture and will lead to a deeper understanding of their mysteries.

\section*{Acknowledgements}
We would like to thank Naritaka Oshita for his collaboration in the initial stage of this project. We are grateful to Alok Laddha for discussion on gravitational memory effects and to Donal O'Connell for conversations. Finally, we are sincerely grateful to Rafael Aoude, Asaad Elkhidir and Matteo Sergola, for useful discussions and feedbacks on earlier versions of the draft. A.C. is also indebted to a for many useful conversations on mass-changing amplitudes and related topics. The work of K.A. was supported by JSPS KAKENHI Grant Nos.~JP24K17046 and JP24KF0153. A.C. was supported by JSPS KAKENHI Grant No.~JP24KF0153. YTH is supported by MoST Grant No. 109-2112-M-002 -020 -MY3 and 112-2811-M-002 -054 -MY2.  


\bibliographystyle{JHEP}
\bibliography{biblio}

\end{document}